# Turbulent Flame Speed Scaling for Expanding Flames with Markstein Diffusion Considerations

Swetaprovo Chaudhuri[1,2], Fujia Wu[1] and Chung K. Law[1,*]

[1]*Department of Mechanical and Aerospace Engineering,*

*Princeton University, Princeton, NJ 08544-5263, USA*

[2]*Department of Aerospace Engineering*

*Indian Institute of Science, Bangalore 560012, India*

**Abstract:** In this paper we clarify the role of Markstein diffusivity, namely the product of the planar laminar flame speed and the Markstein length, on the turbulent flame speed and it's scaling, based on experimental measurements on constant-pressure expanding turbulent flames. Turbulent flame propagation data are presented for premixed flames of mixtures of hydrogen, methane, ethylene, *n*-butane and dimethyl ether with air, in near isotropic turbulence in a dual-chamber, fan-stirred vessel. For each individual fuel/air mixture presented in this work and the recently published $C_8$ data from Leeds, normalized turbulent flame speed data of individual fuel/air mixtures approximately follows a $\operatorname{Re}_{T,f}^{0.5}$ scaling, for which the average radius is the length scale and thermal diffusivity is the transport property of the turbulence Reynolds number. At a given $\operatorname{Re}_{T,f}$, it is experimentally observed that the normalized turbulent flame speed decreases with increasing Markstein number, which could be explained by considering Markstein diffusivity as the leading dissipation mechanism for the large wavenumber, flame surface fluctuations. Consequently, by replacing thermal diffusivity with the Markstein diffusivity in the turbulence Reynolds number, it is found that normalized turbulent flame speeds could be scaled by $\operatorname{Re}_{T,M}^{0.5}$ irrespective of the fuel, equivalence ratio, pressure and turbulence intensity for positive Markstein number flames.

[*]Corresponding author: cklaw@princeton.edu

**I. Introduction**

The turbulent flame speed is an essential parameter in turbulent combustion research, as addressed by the large volume of analytical [1-6], experimental [7-12], computational [13-14] and review literature [15-19]. Assuming that the turbulent flame speed $S_T$ is a meaningful physical quantity, there is the interest to seek a unified scaling description, at least under some special upstream cold flow conditions such as those in isotropic turbulence. Apart from fundamental understanding, such a unified scaling description, if it exists, could for example be utilized as a subgrid scale model for Large Eddy Simulations of combustion processes in engines to supernova explosions.

Under the well-accepted hypothesis of Damköhler that turbulent flame speed is mainly controlled by the total flame surface area, the problem of turbulent flame propagation can be phenomenologically considered to be controlled by two mechanisms, namely production and destruction/dissipation of the flame surface fluctuations. The production might occur due to the following mechanisms: (a) stretching and wrinkling through turbulence, (b) Darrieus-Landau/hydrodynamic instability due to thermal expansion, and (c) diffusional thermal instability for Lewis number ($Le$) < 1 mixtures. The dissipation of flame surface fluctuation, on the other hand, might occur due to: (d) Huygens propagation/kinematic restoration,(e) dissipation at small scales due to thermal conduction and non-equildiffusion caused by $Le > 1$, which will be demonstrated and explained in the sequel, and (f) chemical time scale effects leading to extinction.

The globally-spherical expanding flame is a relatively clean flame configuration and has been frequently used to measure laminar and turbulent flame speeds. Nevertheless, while there is clear relation between an expanding spherical flame and its planar counterpart in the laminar situation, there are two important differences between an expanding turbulent flame and its planar turbulent counterpart. First, similar to expanding laminar flames that are subjected to the curvature induced stretch, expanding turbulent flames are also subjected to a global mean stretch. This could necessitate a correction of the turbulent flame speed as it may not be proportional to the total flame surface area if the mean stretch modifies the local laminar flame speed. Secondly and perhaps more importantly, since the flame is centrally ignited and continuously expands and wrinkled by imposed turbulence, its brush thickness and effective hydrodynamic scales (outer scales) is also increasing while the smallest length scales (inner scales) remain unchanged. This brings an acceleration mechanism to expanding turbulent flames.

The acceleration of expanding turbulent flames was experimentally demonstrated in [20], in which unity *Le* turbulent expanding flames in near isotropic turbulence without hydrodynamic instability was considered, thus eliminating mechanisms (b), (c) considered above. If balance between gas expansion and kinematic restoration is also assumed due to their similar scaling [21], then (d) and *Le*-dependent part of (e) are also eliminated. The following scaling for normalized turbulent flame speed was experimentally arrived at, which includes mechanisms (a) and the *Le*-independent part of (e):

$$\frac{S_T}{S_L} \sim \left[ \frac{u_{rms}}{S_L} \frac{\langle R \rangle}{\delta_L} \right]^{1/2} \quad (1)$$

where $S_L$ is the planar laminar flame speed and $\delta_L$ the corresponding laminar flame thickness, $\langle R \rangle$ is the average radius of the expanding flame and $u_{rms}$ the root-mean-square velocity fluctuations of the cold flow.

Despite of the success of Eqn. (1) on unity *Le* flames, it is widely recognized that *Le* of most mixtures can deviate substantially from unity to affect the flame temperature and subsequently the flame speed through the intrinsic Arrhenius kinetics. Indeed, measurements of laminar flame speeds suffered large scatters until the *Le* effect through flame stretching was recognized in [22]. Furthermore, turbulent flame speed measurements have also shown the influence of *Le* in various studies reviewed in [18]. Therefore it is of interest to seek a scaling relation which takes into account the *Le* effects and is valid for all fuel/oxidizer mixture properties.

The effect of non-unity *Le* in modifying the local flame speed is most prominent in the presence of local stretch and curvature due to the nonequidiffusion of heat and species which are directed normal to the local flame surface. The dependence of the local flame speed on strain and curvature is essentially nonlinear; however, it has been shown by Chen [23] and Kelley *et al*. [24] for the outwardly expanding spherical flame, the empirical linear relation between local flame speed $\tilde{S}_L$ and curvature proposed by Markstein [25] provides a reasonably accurate approximation, *i.e.*,

$$\tilde{S}_L = S_L\left(1-\delta_M \kappa\right) \qquad (2)$$

where $\kappa$ is the curvature with the convex section in the direction of flame propagation being positive, and $\delta_M$ is a coefficient having the unit of length and termed the Markstein length. The value of Markstein length in the unit of flame thickness is the Markstein number, $Mk = \delta_M/\delta_L$, which is a strong function of $Le$ and has different values when measured on the unburned and burned sides of the flame, which we shall designate as $Mk$ and $Mk_b$, respectively, $i.e.$,

$$\tilde{S}_{L,b} = S_{L,b}\left(1-\delta_{M,b} \kappa\right)$$

$$Mk_b = \delta_{M,b}/\delta_L$$

where $S_{L,b}$ is the planar laminar flame speed relative to the burned gas and $\delta_{M,b}$ is the burned gas Markstein length. Using one-step chemistry model an analytical expression for $Mk$ was obtained in [26] and [27], which upon further extension to include temperature-dependent transport properties yields [28],

$$Mk = \frac{\Theta}{\Theta-1}\int_1^\Theta \frac{\lambda}{\lambda_u z}dz + \frac{Ze(Le-1)}{2(\Theta-1)}\int_1^\Theta \frac{\lambda}{\lambda_u z}\ln\left(\frac{z-1}{\Theta-1}\right)dz \qquad (3)$$

where $Ze$ is the Zel'dovich number, $\Theta$ the thermal expansion ratio, $\lambda$ the thermal conductivity, and the subscript $u$ denotes values at the unburned temperature. Although $\Theta$, $Ze$ and $\lambda$ do not greatly vary among different fuel/air mixtures, variations of $Mk$ can still be quite large, primarily due to the variation in $Le$. Hence quantification and fundamental understanding of $Mk$ effects on the turbulent flame speed is a primary necessity for arriving at a unified scaling of the turbulent flame speed, valid for any arbitrary fuel over extensive range of equivalence ratios.

For a turbulent premixed flame with positive Markstein length, the concave portion of the flame, with a negative curvature, propagates faster than the convex portion of the flame which has a positive curvature, leading to reduction of the surface fluctuations and thereby the total flame surface area. This *dissipative* mechanism (e), shown schematically in Figure 1, is most amplified at large wavenumbers or large curvatures. To incorporate such a mechanism into the scaling, statistically planar flames with unity $Le$ and positive $Mk$ was considered theoretically in [29]. The analysis was based on the spectral formulation of the G-equation [30] with turbulence imposing a $k^{-5/3}$ dependence on the flame surface fluctuation spectrum, where $k$ is the wave number. The dependence of the local laminar flame speed on curvature, *i.e.*, Eqn. (2), was incorporated into the G-equation,

$$\frac{\partial G}{\partial t} + \mathbf{V} \cdot \nabla G = \left( S_L - S_L \delta_M \kappa \right) |\nabla G| = S_L |\nabla G| - D_M \kappa |\nabla G| \qquad (4)$$

which essentially provides a Markstein diffusion term with the $D_M = S_L \delta_M$ as the Markstein diffusivity. By further assuming that: (i) balance of dissipation by kinematic restoration and amplification by thermal expansion due to their similar wavenumber dependence (but with opposite signs) [21]; (ii) the dominant role of dissipation by Markstein diffusion at large wavenumbers and thereby retaining Markstein diffusion as the sole dissipation mechanism; and (iii) turbulent flame speed is proportional to the total flame surface area, the following scaling relation (Eqn. (69) in [29]) was arrived:

$$\frac{S_T}{S_L} \sim \left[ \frac{1}{Mk} \frac{u_{rms}}{S_L} \frac{L_I}{\delta_L} \right]^{1/2} \sim \left[ \frac{u_{rms}}{S_L} \frac{L_I}{\delta_M} \right]^{1/2} \qquad (5)$$

where $L_I$ is the integral scale. The unity *Le* assumption of [29] can be easily generalized to any *Le* as long as *Mk* is positive and mean stretch effects on the turbulent flame speed are accounted for.

The present study has two major motivations. First, we will present an extensive experimental database of turbulent flame speeds using expanding turbulent flames. In particular, new data for the following fuels will be examined: hydrogen ($H_2$), ethylene ($C_2H_4$), *n*-butane ($C_4H_{10}$) and dimethylether (DME, $C_2H_6O$), along with the methane ($CH_4$) data of [20] and the iso-octane ($C_7H_{16}$) data of [31], thereby forming a highly diverse group in terms of chemistry as well as distinct, positive *Mk*. Second, based on the experimental data and Eqn. (5), a modified scaling relation for expanding turbulent flames will be demonstrated. We have focused on conditions with positive *Mk*, which are also free from intrinsic flamefront instabilities. It is further noted that all lean premixed flames of large hydrocarbon/air mixtures are characterized by positive *Mk*, implying that the interest for such intrinsic instability-free condition naturally stems from the practical point of view as well.

The final results presented in this paper, concerning scaling of turbulent flame speeds, necessarily involves some empiricism and extrapolation of results from statistically planar flames. It is recognized that in the absence of a complete first principle theory for such conditions - a common problem with most turbulent flows, such empiricism is needed to arrive at useful results.

In Section II, we will first describe the experimental setup. In Section III, the experimental data and results will be presented and discussed. The theoretical consideration between statistically planar flames and expanding flames will be given in Appendix A.

## II. Experimental Setup

### A. Combustion vessel

The experiments were conducted in a nearly constant-pressure apparatus that has been extensively employed in the study of laminar flames [33]. As shown in Figure 2, the apparatus consists of an inner chamber situated within an outer chamber of much larger volume. The inner chamber consists of the cylindrical experimental test section (inner diameter = 114 mm, outer diameter = 165 mm, length = 127 mm), with a near-unity aspect ratio. The two ends of the chamber are sealed with 25 mm thick, 127 mm diameter, quartz windows which allow optical access. The inner chamber is filled with the test combustible gas while the outer chamber with an inert gas of the same density. The two chambers can be opened to each other at the instant of spark ignition by rotating a sleeve that otherwise covers a matrix of holes connecting them, and the propagating flame is automatically quenched upon contacting the inert gas in the outer chamber. The flame propagation event is therefore basically isobaric because of the small volume of the inner chamber relative to that of the outer chamber, hence preventing any significant influence by the global pressure rise on the local flame structure. Another advantage of the design is that experiments can be conducted under high initial pressures, up to 30 bars as in the studies of Refs. [33]-[34], while preserving the integrity of the optical windows. Four fans of 69 mm diameter are located at its walls and are driven by motors situated in the outer chamber. Turbulence is generated by these orthogonally positioned fans as in [7] which continuously run during the entire flame propagation event. The fan-generated, non-reacting turbulent flow field was characterized by high-speed particle image velocimetry (HS-PIV). Detailed flow-field statistics and quantification of the small but unavoidable deviation from isotropy are given in [20]. All turbulent flame speed data were acquired by high-speed schlieren imaging because of its advantages over other methods, for these particular variants of flames and measurements [35]. Furthermore, high-speed Mie scattering images of the expanding flame from a planar laser sheet were obtained to determine the flame brush thickness in support of the choice of scaling parameters. The experiments were conducted at pressures of 1, 2, 3 and 5 atm and with $u_{rms}$ between 1.34 and 5.33 m/s. The domain of interest was chosen to be $0.21 \leq \langle R \rangle / R_{chamber} \leq 0.38$, identified from laminar flame speed experiments to avoid ignition and wall effects at the initial and final stages of flame propagation respectively. Here $R_{chamber} = $ inner diameter/2. This allowed measurements to initiate after an eddy turnover time of $\tau_{ed} = L_I / u_{rms}$, with $L_I \sim 4mm$ for all conditions in the present experiments.

**B. Regime of experimental conditions**

The flame and flow properties as well as the symbols that designate each condition of the experiment in subsequent figures are given in Table 1. For comparison and scaling purpose in Section IV we have also considered the recent experimental data for iso-octane from Leeds [31]. The flame and flow properties of the Leeds data are listed in Table 2. Our fuel matrix constitutes a diverse range in the diffusive-reactive properties not only in terms of molecular diffusivity: from the light $H_2$ (Molecular Weight = 2) to the almost diffusionally neutral $C_2H_4$ (28) relative to the abundant inert species $N_2$, to the relatively heavy hydrocarbon iso-octane (100), but also in terms of chemical reactivity in that each of the fuel species embodies a specific kinetic feature that distinguishes it from the rest. Specifically, relative to $C_4H_{10}$, which can be considered as a typical paraffinic fuel, $CH_4$ is the only *n*-alkane constituted entirely by C-H bonds, rendering the initiation reaction more difficult; $C_2H_4$ is the smallest alkene with a double, C=C, bond and having a high adiabatic flame temperature with correspondingly large laminar flame speeds; iso-octane has a branched structure that renders it knock resistant; and of course $H_2$ has a distinctively different chemistry than the hydrocarbons.

In addition to the above fuels, inclusion of DME adds further diversity to the fuel matrix under consideration in terms of fuel chemistry. DME has recently received considerable attention due to its fairly simple molecular structure yet intricate chemical effects, perhaps the most noteworthy being its NTC (Negative temperature coefficient) behavior. In particular, while the NTC is generally recognized as a low-temperature phenomenon, it could exert a stronger influence at elevated pressures such as those within internal combustion engines and for the present, O(10atm), high-pressure experiments. As such, acquisition of these turbulent flame speed data for further simulation studies employing the DME chemistry is useful in its own right. It is s significant to note from the onset that, in spite of the extensive range of fuel and flow conditions covered herein, the proposed scaling of this work approximately holds for all positive *Mk* irrespective of the fuel/air mixture, fuel chemistry, pressure and turbulence intensity.

Figure 3 plots the conditions of both the present experiments and the Leeds data in the regime diagram proposed in [32]. It is seen that many of the experimental conditions fall in the "thin reaction zone" regime, corresponding to the flame time scale ($t_F = \delta_L / S_L$) > Kolmogorov timescale ($t_\eta$) *i.e.*, $Ka > 1$. However, even in such a regime, the flame structure may not be dominated by turbulent transport such that Lewis and Markstein number effects vanish. This is because eddies of Kolmogorov length scale ($\eta$) might not have sufficient momentum to transfer mass and heat, in and out of the preheat zone. Also, small-scale dissipation intermittency is a well-known phenomenon, which may largely reduce the statistical significance of eddies of Kolmogorov length scale ($\eta$) on the flame structure at moderate turbulence Reynolds numbers. Figure 3 also plots the line $Da = 1$ corresponding to $\delta_L = L_I$, as well as the line $\delta_L = 13\eta$, which is the characteristic length scale located at the centroid of the dissipation spectrum, suggested by Pope [19]. These two lines indicate the boundaries above which turbulent transport is expected to be predominant in the preheat zone. It is seen that the conditions we considered are below these two lines. This indicates that the flamelet structure is at least partially preserved in our experimental conditions.

**III. Results**

In the following, we present turbulent flame speed data from experiments performed with the mixtures in Table 1, *i.e.*, $H_2$/air, $\phi=4.0$; $C_2H_4$-15%, $O_2$-85%, $\phi=1.3$; $C_2H_4$/air, $\phi=1.3$; *n*-$C_4H_{10}$/air, $\phi=0.8$ and $C_2H_6O$/air $\phi=1.0$; $N_2$ is the inert for all the mixtures. In addition, the $CH_4$/air, $\phi=0.9$ data from Ref [20] are also used for comparison. The values of $\delta_{M,b}$ in Table 1 for all conditions were obtained from laminar expanding flame experiments using the approach detailed in [34]. In the rest of this paper any property with a subscript "*b*" indicates measurement with respect to the burned gas while its absence thereof indicates measurement with respect to the unburned gas. For instance, $S_L$ and $\delta_M$ respectively represent the planar laminar flame speed and Markstein length with respect to the unburned gas, while $S_{L,b} = \Theta S_L$ is the planar laminar flame speed with respect to the burned gas.

## A. Raw data

First we plot the raw data of all fuel/air mixtures used in the current study in Figure 4. In Figure 4a $\langle R \rangle$ vs. time is plotted, with $\langle R \rangle$ defined as $\langle R \rangle = \sqrt{A/\pi}$ where $A$ is the area enclosed by the flame edge tracked from the high-speed schlieren imaging. Figure 4b plots $d\langle R \rangle / dt$ vs. $\langle R \rangle$, where $d\langle R \rangle / dt$ is the derivative of $\langle R \rangle$ with respective to time, obtained using a central difference scheme. To avoid ignition and wall effects, only data between $12mm \leq \langle R \rangle \leq 22mm$ are considered. From Figure 4a it is seen that all the $\langle R \rangle$–time curves concave upward, indicating flame acceleration. Figure 4b shows that although the values of $d\langle R \rangle / dt$ for different mixtures vary from 2m/s to 17m/s, they all increase with $\langle R \rangle$. This means that acceleration is observed for all the expanding turbulent flames considered herein.

## B. Power-law dependence of individual experiments

To investigate the magnitude of the flame acceleration, a viable choice is to consider power-law dependence. Here we demonstrate how well the flame speed data of our individual experiments follow the power-law, and the value of the power-law exponent. Figure 5 plots $d\langle R \rangle / dt$ vs. $\langle R \rangle$ for CH$_4$/air mixtures with $\phi=0.9$, $p=1$atm, $u_{rms} = 2.85$m/s. Both the individual data of seven independent runs and their ensemble averages are presented. To obtain the ensemble average of $d\langle R \rangle / dt$ at a common $\langle R \rangle$, they were interpolated with respect to a common $\langle R \rangle$ using the piecewise cubic Hermite interpolation technique (pchip) in Matlab. From Figure 5 it is seen that while the data for individual runs scatter, the ensemble average of $d\langle R \rangle / dt$ clearly increases with $\langle R \rangle$. A least-square fitting of the ensemble averaged data yields $d\langle R \rangle / dt \sim \langle R \rangle^{0.67}$. It is noted that the variation of $\langle R \rangle$ is small for individual experiments, only from 1.2 cm to 2.2 cm, and as such the validity of this power-law dependence for individual experiments is limited. However, as will be shown later, by considering all the experiments in a single scaling, the variation of the dimensionless independent variable, $\langle R \rangle$ scaled by the flame thickness (or Markstein length) is quite large, hence supporting the possible validity of the power law scaling.

**C. Effective turbulence intensity**

Previous studies have attempted to explain the acceleration of expanding turbulent flames as a consequence of the increased turbulence intensity the flame experiences as it expands [7]. Indeed, unlike a statistically stationary flame which experiences constant turbulence intensity, the effective turbulence intensity "experienced" by the expanding flame surface should be different at different stages of expansion. This is because only eddies smaller than the flame size are capable of wrinkling the flame surface, while the larger eddies would convect it like an inertial particle.

In view of this consideration, we have investigated the dependence of effective turbulence intensity on flame size. We consider a quantity

$$u''_{eff} = \left\langle \left( u_r - \langle u_r \rangle_R \right)^2 \right\rangle^{1/2}$$

where $\langle u_r \rangle_R$ is the average of an instantaneous (one-time pdf) realization of $u_r$ over a domain of radius $R$, and $u_r$ is the radial component of the local velocity. Being always positive but a fluctuating quantity, $u''_{eff}$ can be averaged over different realizations (say, averaged in time) to yield $u'_{eff} = \overline{u''_{eff}}$. It is noted that our flow is near isotropic and small deviations from isotropy were quantified in [20]. From our HS-PIV measurements in the non-reacting flow, it was found that in the domain of the reported measurements, and fan speed ranging from 2000 to 8000 rpm, the $u'_{eff}$ can be approximately represented by the following correlation for all the cases under study:
$u'_{eff} = \left( 0.0018875 \times rpm + 0.06622 \right) \times \langle R \rangle^{0.23}$.

Based on the analytical result in [29], *i.e.*, Eqn. (5), this variation in $u'_{eff}$ will introduce another dependence of the turbulent flame speed as $\langle R \rangle^{0.115}$. From here on we will work with $u'_{eff}$ instead of the conventional $u_{rms}$. However, it is important to note that this dependence of $u'_{eff}$ on $\langle R \rangle$ cannot fully explain the power-law dependence of $d\langle R \rangle / dt$ on $\langle R \rangle$ observed in Figure 5.

## D. Best fitting of $d\langle R \rangle / dt$ for all conditions

In order to have a unified understanding of how turbulent flame speed depends on different flow and flame parameters, and achieve a unified scaling of the present results, we consider a surface power-law fitting given by,

$$\frac{1}{S_{L,b}} \frac{d\langle R \rangle}{dt} \propto \left( \frac{u'_{eff}}{S_L} \right)^m \left( \frac{\langle R \rangle}{\delta_{M,b}} \right)^n \tag{6}$$

where *m* and *n* are the fitting parameters. The reason for considering such an expression for scaling is motivated by previous analysis on the turbulent flame speed [29], *i.e.*, Eqn. (5), and previous experimental study on unity *Le* mixtures [20]. We note that the integral scale $L_I$ in Eqn. (5) is replaced by the mean flame size $\langle R \rangle$ in Eqn. (6), which was also done in the scaling presented in [20]. The rationale for such a replacement is that for statistically planar turbulent flames the flame brush thickness is proportional to $L_I$ [32], while for expanding turbulent flames our measurements on flame brush thickness show that it is approximately proportional to the flame size $\langle R \rangle$, as shown in Figure A1. The reason is that as an expanding turbulent flame grows, it experiences the wrinkling of larger and larger flow eddies. A more detailed analysis and measurements on the effects of increase in flame brush thickness are given in Appendix A I.

Figure 6 shows the best surface fit of Eqn. (6) with the experimental data ($C_0$-$C_4$). The best fitting exponents obtained by maximizing $R^2$ are *m* = 0.43 and *n* = 0.45 respectively. The $R^2$ value of the best fitting is 0.92, suggesting the validity of such a scaling. Figure 6 is therefore the main result of this study. In the following, we will provide further discussion on the result and explanation of the scaling, particularly the Markstein diffusion effect.

## IV. Further Discussion and Scaling

## A. Scaling with flame thickness as flame scale

In [20], we have experimentally demonstrated the scaling with flame thickness as the flame scale, *i.e.*, Eqn. (1) for unity *Le* mixtures. Let us now look at how this scaling works for non-unity *Le* mixtures. Figures 7a and b show $S_{L,b}^{-1} d\langle R \rangle / dt$ vs. $\text{Re}_{T,f}^{0.5} = [(u'_{eff}/S_L)(\langle R \rangle/\delta_L)]^{0.5}$ for all the conditions in linear and log-log plots, respectively, and Figure 7c plots $S_{T,c=0.5}/S_L$ vs. $\text{Re}_{T,f}^{0.5}$, with $S_{T,c=0.5}$ calculated from $d\langle R \rangle/dt$ and the thermal expansion ratio based on Eqn. (A10) in Appendix B. It is seen that each set of data corresponding to a *particular* mixture collapses reasonably well on a line, suggesting the general validity of Eqn. (1) for individual fuel/air mixtures (with constant $Mk$). However, for different mixtures the scaling lines have different slopes. Specifically, the slope decreases with increasing $Mk_b$. The C$_2$H$_4$ mixtures (15% O$_2$, 85% N$_2$, $\phi$=1.3), corresponding to negative $Mk_b$, has the largest slope, while the H$_2$/air, $\phi$=4.0 mixture, corresponding to the highest $Mk_b = 4.1$, has the smallest slope. In addition, in Figure 7 we also plot the Leeds's ensemble averaged iso-octane/air data [31]. The ensemble average is calculated based on instantaneous data taken from [31]. A plot of the ensemble averaged data superimposed on the instantaneous data is given in Figure A3. For iso-octane/air, $Mk_b$ decreases from lean to rich and correspondingly $S_{T,c=0.5}/S_L$ increases from lean to rich. Hence, it is seen from Figure 7 that even though most of the flames are in the so-called "thin reaction regime", we see strong influence of molecular diffusion, with $S_{T,c=0.5}/S_L$ varying by more than a factor of 4 for various $Mk_b$.

## B. Scaling with Markstein length as flame scale

Since both the exponents *m* and *n* are close to 0.5 from the result of the best surface fitting based on Eqn. (6), the following scaling is expected to scale flame speeds of all mixtures with positive $Mk_b$,

$$\frac{1}{S_{L,b}} \frac{d\langle R \rangle}{dt} \propto \sqrt{\frac{u'_{eff}}{S_L} \frac{\langle R \rangle}{\delta_{M,b}}} \tag{7}$$

In Figures 8a-c we re-plot the same dataset used in Figure 7 but with $\left[\left(u'_{eff}/S_{L,b}\right)\left(\langle R\rangle/\delta_{M,b}\right)\right]^{1/2}$ as the abscissa. It is seen that the data from various fuels, at different equivalence ratios, pressure and turbulence intensity indeed collapse rather well.

Since identifying the individual fuel/air mixture data is somewhat difficult in the data band, in Figure 8d only the two most widely tested fuel/air mixtures in the current study: DME/air with $\phi=1.0$ and ethylene/air with $\phi=1.3$ are plotted. Clearly they collapse on a narrow and apparently linear band despite their different chemistry, laminar flame speeds and Markstein lengths.

**C. Markstein diffusion on flame morphology**

Finally, we can also qualitatively appreciate the importance of Markstein diffusion in turbulent flame propagation by examining the flame morphology. We note from Eqn. (7) that the Markstein diffusivity $D_M = S_L \delta_M$ is the only parameter concerning the properties of the mixture in these equations. With the flow parameters same, Markstein diffusivity plays a critical role in the dissipation of wrinkles on the flame surface.

To demonstrate this, we plot in Figures 9a-b the schlieren images of $CH_4$/air, $\phi=0.9$ flames at 5 atm pressure and fan speeds of 2000 and 4000 rpm, and in Figures 9c, d the images of $H_2$/air, $\phi=4.0$ flames at a pressure of 5 atm and fan speeds of 4000 and 7500 rpm. Comparing the two $CH_4$/air flames with the same Markstein length, it is seen that the flame at 4000 rpm shows evolution of finer scale structures as compared to the flame at 2000 rpm, which is due to the increased turbulence intensity. However, at the same and even higher turbulence intensities, we see the smallest scale structure for the two $H_2$/air flames is much larger than that of the $CH_4$/air flames. This cannot be explained by the flame thickness effect because $\delta_L$ of the $H_2$/air flames is about half of that of $CH_4$/air flames. Rather, it can be explained by the Markstein length or Markstein diffusivity: due to the high reactivity and large Lewis number of the rich $H_2$/air flames, both $S_{L,b}$ and $\delta_{M,b}$ are much higher than those of $CH_4$/air flames, leading to higher $D_{M,b}$. The large Markstein diffusivity prevents evolution of fine scale structures for the $H_2$/air flames due to its inherent dissipative property at small scales, but a smaller Markstein diffusivity allows finer scale flame wrinkling for the $CH_4$/air flames. Indeed, if we compare

Figures 9a and 9d, which have nearly the same values of $\text{Re}_{T,f} = u'_{eff} \cdot \langle R \rangle / (S_L \cdot \delta_L)$, 1287 and 1267, the flame in the former image shows much finer scales in comparison to the one in the latter. This means that $\text{Re}_{T,f}$ is not the governing parameter. However, if one considers $\text{Re}_{T,M} = u'_{eff} \cdot \langle R \rangle / (S_L \cdot \delta_M)$ as in Eqn. (7) as the relevant turbulence Reynolds number, $\text{Re}_{T,M}$ of Figure 9d, 269, is much smaller than that of Figure 9a, 798, which readily explains the difference in the smallest scale shown in the images.

## V. Conclusions

We have presented experimental turbulent flame speed data for $H_2$, $CH_4$, $C_2H_4$, $C_2H_6O$ and $n$-$C_4H_{10}$ mixtures with air, obtained in the present study, as well as the data for $iso$-$C_8H_{18}$ from [31]. Using such a large database comprising over various fuels at large ranges of turbulence intensity and pressure, we have shown that irrespective of the fuel, the normalized turbulent flame speed data follows the $\text{Re}_{T,M}^{0.5}$ scaling, in which the average radius is the length scale and Markstein diffusivity is the transport property. Overall, the final result of Figures 8a-d indicates the possibility of a unified, fuel-invariant scaling of the normalized turbulent flame speed. For moderate to large $\text{Re}_T$, the proposed scaling of Eqn. (7) is apparently and approximately valid for expanding flames.

**Acknowledgements:**

This work at Princeton University was supported by the Air Force Office of Scientific Research under the technical monitoring of Dr. Chiping Li. SC continued his participation subsequent to joining his present affiliation – the Indian Institute of Science. The authors sincerely acknowledge the valuable comments of Dr. Alan R. Kerstein, Dr. Andrei Lipatnikov, as well as an anonymous reviewer.

**Appendix A**:

**Expanding turbulent flames vs. statistically planar turbulent flames**

**I. Increase of effective hydrodynamic length scale**

For statistically planar flame propagating in homogenous isotropic turbulence, the normalized turbulent flame speed $S_T/S_L$ is assumed to be dominated by the flame surface fluctuation, *i.e.*, Eqn. (56) in [29],

$$S_T/S_L \propto \left\langle \sqrt{1+\nabla g \cdot \nabla g} \right\rangle \sim \sqrt{1+\int_{k_I}^{\infty} k^2 \Gamma(k;u_{rms},L_H,S_L,\delta_M)dk} \quad (A1)$$

where $g = (G-z)$ is the flame surface fluctuation, $G$ the level set function, $\Gamma$ the $g^2$ spectrum, $k_I = 2\pi/L_I$ and $L_I$ the integral scale. The spectrum $\Gamma$ is given by Peters [30] by adopting gradient transport assumption and dimensional consistency,

$$\Gamma(k) = Bk^{-5/3} \exp\left[-3c_1(2\pi)^{1/3}\left(\frac{u_{rms}}{S_L}\right)^{-1}\left(\frac{k}{k_I}\right)^{1/3}\right] \times$$

$$\exp\left[-\tfrac{3}{4}(2\pi)^{4/3}c_2\left(\frac{u_{rms}}{S_L}\frac{L_I}{\delta_M}\right)^{-1}\left(\frac{k}{k_I}\right)^{4/3}\right] \quad (A2)$$

where $B, c_1, c_2$ are constants. This expression includes the fluctuation production term $k^{-5/3}$ imposed by the isotropic turbulence, and two exponentially decaying dissipation terms. The first dissipation term is due to the nonlinear Huygens propagation or kinematic restoration, while the second dissipation term is due to the dissipation by Markstein diffusion (with positive $Mk$). We note that this expression for $\Gamma(k)$ includes all the energy at all scales smaller than $L_I$. In [29] the dissipation term by kinematic restoration in Eqn. (A2) is further neglected and Eqn. (A1) can be analytically integrated, yielding Eqn. (5), which is also Eqn. (69) in [29], *i.e.*,

$$\frac{S_T}{S_L} \propto \left[\frac{1}{Mk}\frac{u_{rms}}{S_L}\frac{L_I}{\delta_L}\right]^{1/2} \propto \left[\frac{u_{rms}}{S_L}\frac{L_I}{\delta_M}\right]^{1/2} \quad (A3)$$

A similar form of turbulent flame speed scaling: i.e. $\delta_L$ and not $\delta_M$ was also obtained by Damköhler [41] in the limit of small-scale turbulence. The key step to arrive at a scaling for $S_T/S_L$ is by assuming equality of the chemical time scale in the laminar and turbulent cases. Therefore $S_T/S_L \sim \sqrt{D_T/D_{th}}$, where $D_T$ is the turbulent diffusivity. Dimensional arguments suggest $D_T$ should have units of $[L]^2[T]^{-1}$ and hence it is conventionally scaled as $D_T = u_{rms} L_I$ for non-reacting statistically stationary flows.

An important assumption made both in [29] and in Damköhler's derivation is that the effective hydrodynamic scale of the flame is equal to the integral scale $L_I$. In [32] and [20] the flame brush thickness was assumed to be equal to $L_I$.

$$\delta_T = \langle g^2 \rangle^{1/2} = L_I \tag{A4}$$

For a statistically planar stationary flame considered in [29], the flame brush thickness is a steady quantity (after the initial transience) and should be controlled by the large-scale eddies of turbulence. Hence this was a plausible assumption. In fact this was the main reason behind the appearance of $L_I$ in the turbulent flame speed formula. However, for an expanding flame Eqn. (A4) is not necessarily appropriate. As an expanding turbulent flame grows in its size, its effective hydrodynamic length must also increase. This is especially true for the cases in the present experiments where the size of the flame is instantaneously much smaller than the domain size.

To confirm the increase of flame brush thickness, we performed analysis of the Mie scattering images of expanding flames for some of the fuel/air mixtures in Table 1. We estimate the flame brush thickness based on the standard derivation of the fluctuating flame radius. In Figure A1, the standard deviation of the flame radius obtained from Mie scattering images is plotted with respect to the mean radius. Each of the several experimental conditions was repeated at least six times. From the measurements, we found the standard derivation of the flame radius is approximately proportional to the mean flame radius, and based on a 96% probable confidence interval, we estimate the proportionality is approximately $\delta_T \approx \langle R \rangle / 2$.

Consequently, a plausible assumption of the expanding turbulent flame is the effective hydrodynamic scale and the flame brush thickness is proportional to the flame radius, *i.e.*,

$$\delta_T \propto \langle R \rangle \propto L_{\text{expanding flame}} \tag{A5}$$

Furthermore, to arrive at a reasonable scaling for expanding turbulent flames, the integral scale $L_I$ in Eqns. A1-A3 should be replaced by $\delta_T$ or $\langle R \rangle$, yielding

$$\frac{S_T}{S_L} \propto \left( \frac{u_{rms}}{S_L} \frac{\delta_T}{\delta_M} \right)^{\frac{1}{2}} \propto \left( \frac{u_{rms}}{S_L} \frac{\langle R \rangle}{\delta_M} \right)^{\frac{1}{2}} \tag{A6}$$

For the same argument, since the effective hydrodynamic scale for expanding flames is increasing and scales with the flame brush thickness and mean flame radius, the effective turbulent diffusivity for the expanding turbulent flame could be considered as $D_T \propto u_{rms} \delta_T \propto u_{rms} \langle R \rangle$. Therefore, Eqn. (A6) can also be attained following Damköhler's derivation.

## II. Effect of global mean stretch

Similar to laminar expanding flames, turbulent expanding flames are also subjected to a global mean stretch, $(2/\langle R \rangle) d\langle R \rangle / dt$. This could cause discrepancy with statistically planar flame results if the local laminar flame speed varies significantly with stretch according to $Mk$, hence weakening the assumption that the turbulent flame speed is proportional to the flame surface area. In general, the local laminar flame speed depends on curvature and strain. According to Kerstein *et al*. [3]:

$$S_T \approx \langle S_L |\nabla G| \rangle = S_L \langle |\nabla G| \rangle - S_L \delta_M \langle \kappa |\nabla G| \rangle - \delta_M \langle s |\nabla G| \rangle \tag{A7}$$

where $s$ is the strain. The second and third terms on the RHS concern with the joint distribution of curvature and strain with the flame surface area ratio respectively which in general are highly non-trivial. The problem could be circumvented by the assumption of a quasi-symmetric distribution of curvature and strain rate although with non-zero means. Then, these two terms can be respectively decomposed as:

$$\langle \kappa |\nabla G| \rangle = \langle -\kappa_n |\nabla G|_n - \kappa_{n-1} |\nabla G|_{n-1} \ldots \ldots \kappa_0 |\nabla G|_0 \ldots \ldots + \kappa_{n-1} |\nabla G|_{n-1} + \kappa_n |\nabla G|_n \rangle \text{ and}$$

$$\langle s |\nabla G| \rangle = \langle -s_n |\nabla G|_n - s_{n-1} |\nabla G|_{n-1} \ldots \ldots s_0 |\nabla G|_0 \ldots \ldots + s_{n-1} |\nabla G|_{n-1} + s_n |\nabla G|_n \rangle$$

where the subscript $n$ represents the $n^{th}$ bin of the positive or negative side of the joint pdf. Clearly if the curvature distribution is nearly symmetric, the negative part is cancelled by the positive part of the distribution since $|\nabla G|_n$ is always positive and independent of the sign of $\kappa$ or $s$. Therefore to the leading order the only terms left would be the product of the mean curvature, mean strain rate and the mean flame surface area ratio respectively. Hence, $S_L \delta_M \langle \kappa |\nabla G| \rangle + \delta_M \langle s |\nabla G| \rangle \sim S_L \delta_M \langle \kappa \rangle \langle |\nabla G| \rangle + \delta_M \langle s \rangle \langle |\nabla G| \rangle$. Thus for the expanding flame for any positive $Mk$, Eqn. (7) can be written as

$$\left( I_{o,b} S_{L,b} \right)^{-1} d\langle R \rangle / dt \sim \left[ \left( u_{eff} / S_L \right) \left( \langle R \rangle / \delta_M \right) \right]^{1/2} \tag{A8}$$

The prefactor is the stretch factor and will be represented by

$$I_{0,b} = 1 - 2\delta_{M,b} / \langle R \rangle - \delta_{M,b} \langle s \rangle / S_{L,b} \tag{A9}$$

In this paper, we have some limited measurement of $\langle s \rangle$ which is found be much smaller than the $2/\langle R \rangle$ term and as such is not considered henceforth. To investigate the effect of this correction term on the normalized turbulent flame speed, in Figure A2, we compare $S_{L,b}^{-1} d\langle R \rangle / dt$ and $\left( I_{0,b} S_{L,b} \right)^{-1} d\langle R \rangle / dt$ with $I_{0,b} = 1 - 2\delta_{M,b} / \langle R \rangle$. It is seen that the effect is very small.

**Appendix B**

**Relation between $d\langle R \rangle / dt$ and $S_T$**

It is noted that $S_{L,b}^{-1} d\langle R \rangle / dt \neq S_T / S_L$ due to gas expansion effects. $S_T / S_L$ could be obtained by the conversion $S_{L,b}^{-1} d\langle R \rangle / dt = c_0 \left( S_T / S_L \right)$ where $c_0$ is a constant for individual mixtures, being weakly dependent on the density ratio $\Theta = \rho_u / \rho_b$. The exact value of this constant is best determined by simultaneous Schlieren and Mie scattering imaging. Bradley *et al.* [8] showed that the ratio $\langle R \rangle / \langle R \rangle_{c=0.5} \sim 1.33 = 4/3$. Here $\langle R \rangle$ is the averaged radius based on the area enclosed by the schlieren edge and $\langle R \rangle_{c=0.5}$ is the average radius at the location where the mean progress variable is $c = 0.5$. Hence, as shown in [20] accounting for gas expansion within the flame brush, the normalized turbulent flame speed at $c = 0.5$ is given by:

$$S_{T,c=0.5}/S_L = \left[2\Theta/(\Theta+1)\right]\left(\langle R\rangle/\langle R\rangle_{c=0.5}\right)^2 \left(S_{L,b}^{-1} d\langle R\rangle/dt\right)$$
$$= \frac{16}{9}\left[2\Theta/(\Theta+1)\right]\left(S_{L,b}^{-1} d\langle R\rangle/dt\right) \quad . \tag{A10}$$

**References:**

[1]    P. Clavin, F. A. Williams, *Journal of Fluid Mechanics* 90 (1979) 589.

[2]    V. Yakhot, *Combustion Science and Technology* 60 (1988) 191.

[3]    A. R. Kerstein, W. T. Ashurst, F. A. Williams, *Physical Review* A 37 (1988) 2728.

[4]    A. R. Kerstein, W. T. Ashurst, *Physical Review Lett*ers 68 (1992) 934.

[5]    A. Lipatnikov and J. Chomiak, *Proceedings of the Combustion Institute* 31 (2007) 1361.

[6]    H. Kolla, N. Rogerson, N. Swaminathan, *Combustion Science & Technology* 182 (2010) 284.

[7]    R.G. Abdel-Gayed, D. Bradley, M. Lawes, *Proc. Roy. Soc. Lond.* A 414 (1987) 389

[8]    D. Bradley, M. Lawes, M.S. Mansour, *Combustion and Flame* 158 (2011) 123

[9]    S. A. Filatyev, J. F. Driscoll, C. D. Carter, J. M. Donbar, *Combustion and Flame* 141 (2005) 1

[10]   H. Kobayashi, T. Tamura, K. Maruta, T. Niioka, *Proceedings of the Combustion Institute* 26 (1996) 389.

[11]   H. Kobayashi, K. Seyama, H. Hagiwara, Y. Ogami, *Proceedings of the Combustion Institute* 30 (2005) 827.

[12]   P. Venkateswaran, A. Marshall, D. H. Shin, D. Noble, J. Seitzman, T. Lieuwen, *Combustion and Flame*, 158 (2011) 1602.

[13]   J.B. Bell, M.S. Day, I.G. Shepherd, M.R. Johnson, R.K. Cheng, J,F. Grcar, et al. *Proceedings of the National Academy of Sciences* 2005;102(29):10006–11.

[14]   Y. Shim, S. Tanaka, M. Tanahashi, T. Miyauchi, *Proceedings of the Combustion Institute* 33 (2010) 1455.

[15]   N. Peters, *Turbulent Combustion*, Cambridge University Press, NY, 2000.

[16]   J. F. Driscoll, *Progress in Energy & Combustion Science* 34 (2008) 91.

[17]   A. N. Lipatnikov, J. Chomiak, *Progress in Energy & Combustion Science* 36 (2002) 1.

[18]   A. N. Lipatnikov, J. Chomiak, *Progress in Energy & Combustion Science* 28 (2005) 1.

[19]  S.B. Pope, *Annual Review of Fluid Mechanics,* 19 (1987), 237

[20]  S. Chaudhuri, F. Wu, D. Zhu, C.K. Law, *Physical Review Letters*, 108, (2012), 044503

[21]  N. Peters, H. Wenzel, F. A. Williams, *Proceedings of the Combustion Institute* 28 (2000) 235.

[22]  C.K. Wu and C. K. Law .*Proceedings of the Combustion Institute,* 20, (1985), 1941.

[23]  Z. Chen, *Combustion and Flame*, 158, (2011), 2, 291.

[24]  A.P. Kelley, J.C. Bechtold and C.K. Law, *Journal of Fluid Mechanics*, 691, (2012), 26.

[25]  G. H. Markstein, Journal of Aeronautical Sciences 18, (1951), 199.

[26]  P. Pelce, P. Clavin, *Journal of Fluid Mechanics* 124 (1982) 219.

[27]  M. Matalon, B. J. Matkowsky, *Journal of Fluid Mechanics* 124 (1982) 239.

[28]  P. Clavin, P. Garcia, *Journal de Mecaniqiue* 2 (1983) 245.

[29]  S. Chaudhuri, V. Akkerman, C.K. Law, *Physical Review E,* 84, (2011) 026322.

[30]  N. Peters,*Journal of Fluid Mechanics* 242 (1992) 611.

[31]  M. Lawes, M. P. Ormsby, C. G.W. Sheppard, R. Woolley, *Combustion and Flame*, 159 (2012) 1949.

[32]   N. Peters, *Journal of Fluid Mech*anics 384 (1999) 107.

[33]   G. Rozenchan, D.L. Zhu, C.K. Law, *Proceedings of the Combustion Institute*, 29, (2002) 1461

[34]  A. P. Kelley and C. K. Law, *Combustion and Flame*, 156, 9, (2009) 1844.

[35]  D. Bradley, M.Z. Haq, R.A. Hicks, T. Kitagawa, M. Lawes, C.G.W. Sheppard, R. Woolley, Combustion and Flame 133 (2003) 415

[36]  M. Fairweather, M.P. Ormsby, C.G.W. Sheppard, R. Woolley, *Combustion and Flame*,156 (2009) 780.

[37]  R. J. Kee, F. M. Rupley, J. A. Miller, et al., CHEMKIN Collection, Release 3.6, 2000.

[38]  H. Wang, X.Q. You, A.V. Joshi, S.G. Davis, A. Laskin, F. Egolfopoulos, C.K. Law, USC Mech Version II. High-Temperature Combustion Reaction Model of $H_2$/CO/$C_1$–$C_4$Compounds, May 2007. <http://ignis.usc.edu/USC_Mech_II.htm>.

[39]  M.P. Burke, M. Chaos, Y. Ju, F.L. Dryer, and S.J. Klippenstein, International Journal of Chemical Kinetics, 44 (2011), 444.

[40]  A.P. Kelley, Dynamics of Expanding Flames, Ph.D. Thesis, Princeton University, 2011.

[41]   G. Damkohler, Z. Elektrochem. 46, 601 (1940) [English Translation: NASA Tech. Mem. 1112 (1947)].

[42]   M. Chaos, A. Kazakov, Z. Zhao, F.L. Dryer, Int. J. Chem. Kinetics 39 (2007) 399–414.

**FIGURES AND TABLES:**

| Symbol | Mixture | φ | p/p$_0$ | rpm | u$_{rms}$ (m/s) | S$_L$ (m/s)[1] | S$_{L,b}$ (m/s)[1] | δ$_L$ (m)[1] | δ$_{M,b}$ (m)[2] | S$_{Lb}$*δ$_L$ | S$_{L,b}$*δ$_{M,b}$ | Mk$_b$ | Mk$_u$ | Θ |
|---|---|---|---|---|---|---|---|---|---|---|---|---|---|---|
| ☆ | H2-air | 4.00 | 5 | 2000 | 1.43 | 1.13 | 5.45 | 8.55E-05 | 3.50E-04 | 4.66E-04 | 1.91E-03 | 4.10 | 5.67 | 4.82 |
| ☆ | H2-air | 4.00 | 5 | 4000 | 2.85 | 1.13 | 5.45 | 8.55E-05 | 3.50E-04 | 4.66E-04 | 1.91E-03 | 4.10 | 5.67 | 4.82 |
| ☆ | H2-air | 4.00 | 5 | 7500 | 5.33 | 1.13 | 5.45 | 8.55E-05 | 3.50E-04 | 4.66E-04 | 1.91E-03 | 4.10 | 5.67 | 4.82 |
| ○ | CH$_4$-air | 0.90 | 1 | 2000 | 1.43 | 0.32 | 2.31 | 4.80E-04 | 5.64E-04 | 1.11E-03 | 1.30E-03 | 1.18 | 3.15 | 7.17 |
| ○ | CH$_4$-air | 0.90 | 1 | 4000 | 2.85 | 0.32 | 2.31 | 4.80E-04 | 5.64E-04 | 1.11E-03 | 1.30E-03 | 1.18 | 3.15 | 7.17 |
| ○ | CH$_4$-air | 0.90 | 2 | 4000 | 2.85 | 0.25 | 1.83 | 2.90E-04 | 3.08E-04 | 5.31E-04 | 5.65E-04 | 1.06* | 3.05 | 7.32 |
| ○ | CH$_4$-air | 0.90 | 3 | 4000 | 2.85 | 0.22 | 1.57 | 2.20E-04 | 2.09E-04 | 3.45E-04 | 3.29E-04 | 0.95* | 2.92 | 7.14 |
| ○ | CH$_4$-air | 0.90 | 5 | 2000 | 1.43 | 0.17 | 1.22 | 1.60E-04 | 1.17E-04 | 1.96E-04 | 1.43E-04 | 0.73 | 2.70 | 7.20 |
| ○ | CH$_4$-air | 0.90 | 5 | 4000 | 2.85 | 0.17 | 1.22 | 1.60E-04 | 1.17E-04 | 1.96E-04 | 1.43E-04 | 0.73 | 2.70 | 7.20 |
| △ | C$_2$H$_4$-15%O$_2$-85%N$_2$ | 1.30 | 2 | 2000 | 1.43 | 0.18 | 1.15 | 3.67E-04 | -4.04E-04 | 4.22E-04 | -4.64E-04 | -1.10 | 0.75 | 6.39 |
| △ | C$_2$H$_4$-15%O$_2$-85%N$_2$ | 1.30 | 2 | 4000 | 2.85 | 0.18 | 1.15 | 3.67E-04 | -4.04E-04 | 4.22E-04 | -4.64E-04 | -1.10 | 0.75 | 6.39 |
| △ | C$_2$H$_4$-15%O$_2$-85%N$_2$ | 1.30 | 5 | 2000 | 1.43 | 0.11 | 0.75 | 2.17E-04 | -1.80E-04 | 1.63E-04 | -1.35E-04 | -0.83 | 1.09 | 6.82 |
| △ | C$_2$H$_4$-15%O$_2$-85%N$_2$ | 1.30 | 5 | 4000 | 2.85 | 0.11 | 0.75 | 2.17E-04 | -1.80E-04 | 1.63E-04 | -1.35E-04 | -0.83 | 1.09 | 6.82 |
| ▽ | C$_2$H$_4$-air | 1.30 | 1 | 2000 | 1.43 | 0.61 | 5.01 | 2.92E-04 | 4.87E-04 | 1.46E-03 | 2.44E-03 | 1.67 | 3.77 | 8.22 |
| ▽ | C$_2$H$_4$-air | 1.30 | 1 | 4000 | 2.85 | 0.61 | 5.01 | 2.92E-04 | 4.87E-04 | 1.46E-03 | 2.44E-03 | 1.67 | 3.77 | 8.22 |
| ▽ | C$_2$H$_4$-air | 1.30 | 2 | 4000 | 2.85 | 0.53 | 4.36 | 1.61E-04 | 2.77E-04 | 7.02E-04 | 1.21E-03 | 1.72 | 3.83 | 8.23 |
| ▽ | C$_2$H$_4$-air | 1.30 | 3 | 2000 | 1.43 | 0.48 | 3.95 | 1.16E-04 | 2.06E-04 | 4.59E-04 | 8.12E-04 | 1.77* | 3.88 | 8.25 |
| ▽ | C$_2$H$_4$-air | 1.30 | 5 | 4000 | 2.85 | 0.41 | 3.42 | 7.85E-05 | 1.47E-04 | 2.68E-04 | 5.03E-04 | 1.87 | 3.98 | 8.26 |
| ◇ | nC$_4$H$_{10}$-air | 0.80 | 5 | 4000 | 2.85 | 0.18 | 1.18 | 1.27E-04 | 3.49E-04 | 1.50E-04 | 4.12E-04 | 2.75 | 4.63 | 6.56 |
| □ | C$_2$H$_6$O-air | 1.00 | 1 | 2000 | 1.43 | 0.47 | 3.92 | 3.24E-04 | 8.71E-04 | 1.27E-03 | 3.41E-03 | 2.69 | 4.81 | 8.33 |
| □ | C$_2$H$_6$O-air | 1.00 | 1 | 4000 | 2.85 | 0.47 | 3.92 | 3.24E-04 | 8.71E-04 | 1.27E-03 | 3.41E-03 | 2.69 | 4.81 | 8.33 |
| □ | C$_2$H$_6$O-air | 1.00 | 5 | 2000 | 1.43 | 0.29 | 2.44 | 9.70E-05 | 2.60E-04 | 2.37E-04 | 6.35E-04 | 2.68 | 4.81 | 8.42 |
| □ | C$_2$H$_6$O-air | 1.00 | 5 | 4000 | 2.85 | 0.29 | 2.44 | 9.70E-05 | 2.60E-04 | 2.37E-04 | 6.35E-04 | 2.68 | 4.81 | 8.42 |
| □ | C$_2$H$_6$O-air | 1.00 | 5 | 6000 | 4.29 | 0.29 | 2.44 | 9.70E-05 | 2.60E-04 | 2.37E-04 | 6.35E-04 | 2.68 | 4.81 | 8.42 |
| □ | C$_2$H$_6$O-air | 1.00 | 10 | 4000 | 2.85 | 0.23 | 1.95 | 5.97E-05 | 6.95E-09 | 1.16E-04 | 1.35E-08 | 2.78 | 4.92 | 8.47 |

[1] Computed with PREMIX in Chemkin II with USC Mech II, PRF Mechanism, and Burke et al. H$_2$ Mechanism Refs [37,38,39]
[2] Data from new and past laminar flame speed experiments [40]; * linearly interpolated between pressures

**Table 1:** Legends, laminar flame properties and turbulent intensity for all present (C$_0$-C$_4$) experimental conditions. $S_L$ and $\delta_L$ are obtained from Premix calculations within Chemkin II. The flame thickness is obtained from the temperature profile by $\delta_L = (T_b - T_u)/(dT/dx|_{max})$. The burnt Markstein length $\delta_{M,b}$ is obtained from laminar spherical flame experiments with the help of nonlinear extrapolation [40].

[1]Computed with PREMIX in Chemkin II with USC Mech II, PRF Mechanism, and Burke et al. H2 Mechanism Refs. [37, 38, 39].

[2]Data from new and past laminar flame speed experiments [40]; *linearly interpolated between pressures.

| Symbol | $\phi$ | $p/p_0$ | $\Theta$ | $S_L$ (m/s)[1] | $S_{L,b}$ (m/s)[1] | $\delta_{M,b}$ (m)[2] | $\delta_L$ (m)[1] | Le | $Mk_b$ | $Mk_u$ |
|---|---|---|---|---|---|---|---|---|---|---|
| • | 0.80 | 1.00 | 6.09 | 0.31 | 1.89 | 1.42E-03 | 4.27E-04 | 2.98 | 3.33 | 5.13 |
| • | 1.00 | 1.00 | 6.84 | 0.43 | 2.94 | 1.13E-03 | 3.72E-04 | 1.43 | 3.04 | 4.96 |
| • | 1.20 | 1.00 | 6.96 | 0.41 | 2.85 | 6.00E-04 | 3.83E-04 | 0.93 | 1.57 | 3.51 |
| • | 1.40 | 1.00 | 6.80 | 0.25 | 1.73 | -8.00E-04 | 5.81E-04 | 0.90 | -1.38 | 0.54 |
| • | 0.80 | 5.00 | 6.11 | 0.21 | 1.28 | 5.20E-04 | 1.31E-04 | 2.98 | 3.96 | 5.77 |
| • | 0.90 | 5.00 | 6.59 | 0.25 | 1.65 | 3.55E-04 | 1.17E-04 | 2.94 | 3.03 | 4.92 |
| • | 1.00 | 5.00 | 6.95 | 0.28 | 1.95 | 3.10E-04 | 1.12E-04 | 1.43 | 2.77 | 4.71 |
| • | 1.10 | 5.00 | 7.05 | 0.29 | 2.04 | 2.60E-04 | 1.11E-04 | 0.94 | 2.33 | 4.29 |
| • | 1.20 | 5.00 | 6.99 | 0.28 | 1.96 | 2.30E-04 | 1.23E-04 | 0.93 | 1.88 | 3.82 |
| • | 1.40 | 5.00 | 6.80 | 0.24 | 1.63 | NA | 2.29E-04 | 0.90 | NA | NA |
| • | 1.60 | 5.00 | 6.60 | 0.15 | 0.99 | NA | 3.61E-04 | 0.88 | NA | NA |
| • | 1.80 | 5.00 | 6.39 | 0.07 | 0.47 | NA | 5.08E-04 | 0.86 | NA | NA |
| • | 2.00 | 5.00 | 6.16 | 0.05 | 0.31 | NA | 6.62E-04 | 0.84 | NA | NA |
| • | 0.80 | 10.00 | 6.12 | 0.16 | 0.98 | 3.10E-04 | 8.19E-05 | 2.98 | 3.78 | 5.60 |
| • | 1.00 | 10.00 | 6.99 | 0.23 | 1.61 | 1.73E-04 | 6.93E-05 | 1.43 | 2.49 | 4.44 |
| • | 1.20 | 10.00 | 7.00 | 0.24 | 1.68 | NA | 8.14E-05 | 0.93 | NA | NA |
| • | 1.40 | 10.00 | 6.81 | 0.21 | 1.43 | NA | 1.48E-04 | 0.90 | NA | NA |

[1] Computed with PREMIX in Chemkin II with PRF Mech, Refs [37,38,42]
[2] Data from laminar flame speed experiments from A. Kelley PhD Thesis: Fig. 6.18 [40]

**Table 2:** Legends, laminar flame properties and turbulent intensity for all experimental conditions of iso-octane (i-$C_8H_{18}$) data from Leeds [31]. The $S_{L,b}$ and $\delta_{M,b}$ are obtained from experiments in our group such that the nonlinear extrapolation technique used for other data sets in Table 1 is consistent throughout. Same colors are assigned for the different $u_{rms}$ cases corresponding to same $\phi$ and $p$ as $u_{rms}$ variation can be identified from $\text{Re}_T$, different variants of which are used as abscissa of most of the figures.

[1]Computed with PREMIX in Chemkin II with PRF Mech, Refs. [37, 38, 42].
[2]Data from laminar flame speed experiments from A. Kelley PhD Thesis: Fig. 6.18 [40]

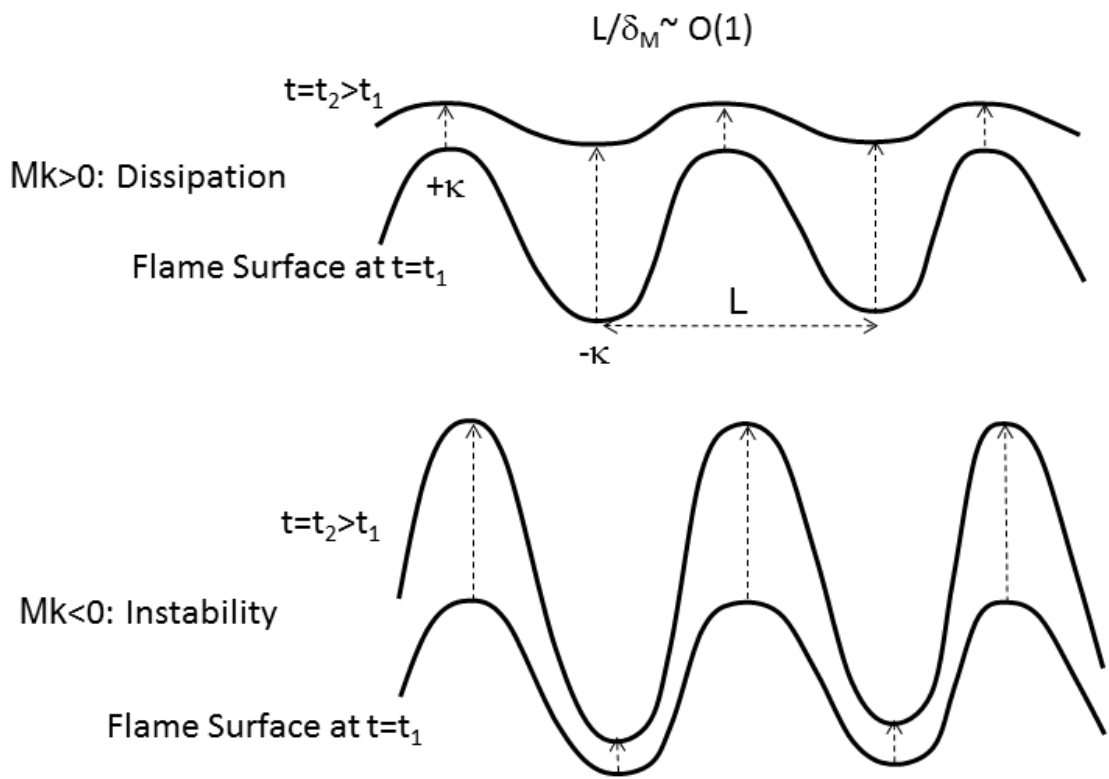

**Figure 1:** Schematic of flame surface fluctuation dissipation and amplification process by positive and negative Markstein numbers (Mk) respectively.

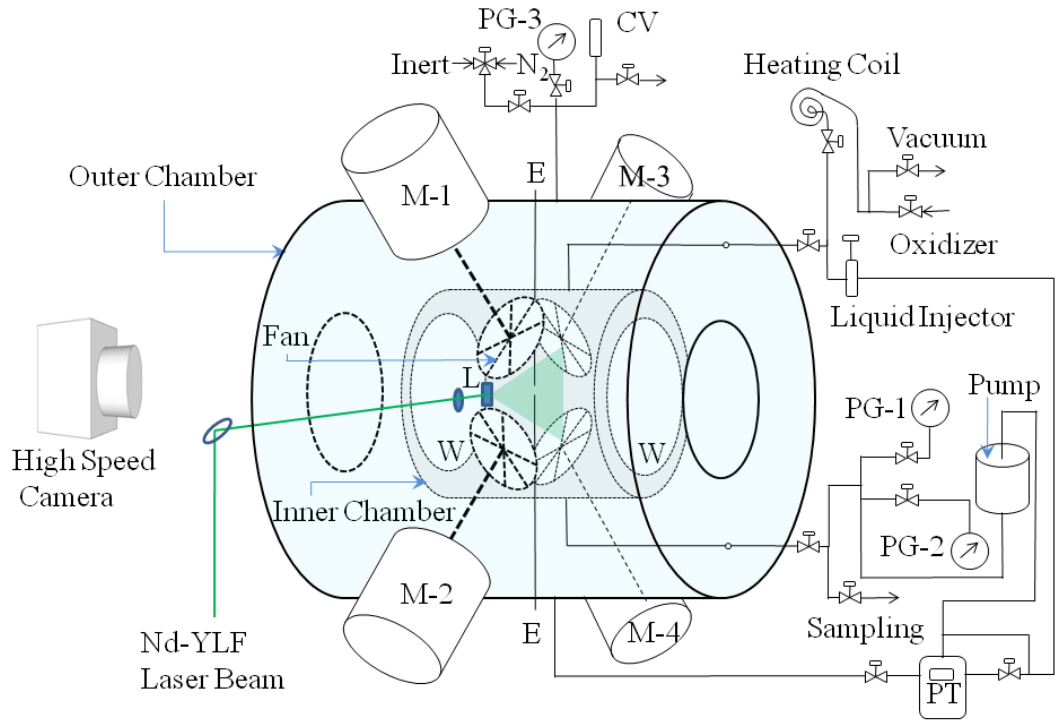

**Figure 2:** Schematic of experimental setup showing the dual chamber apparatus.

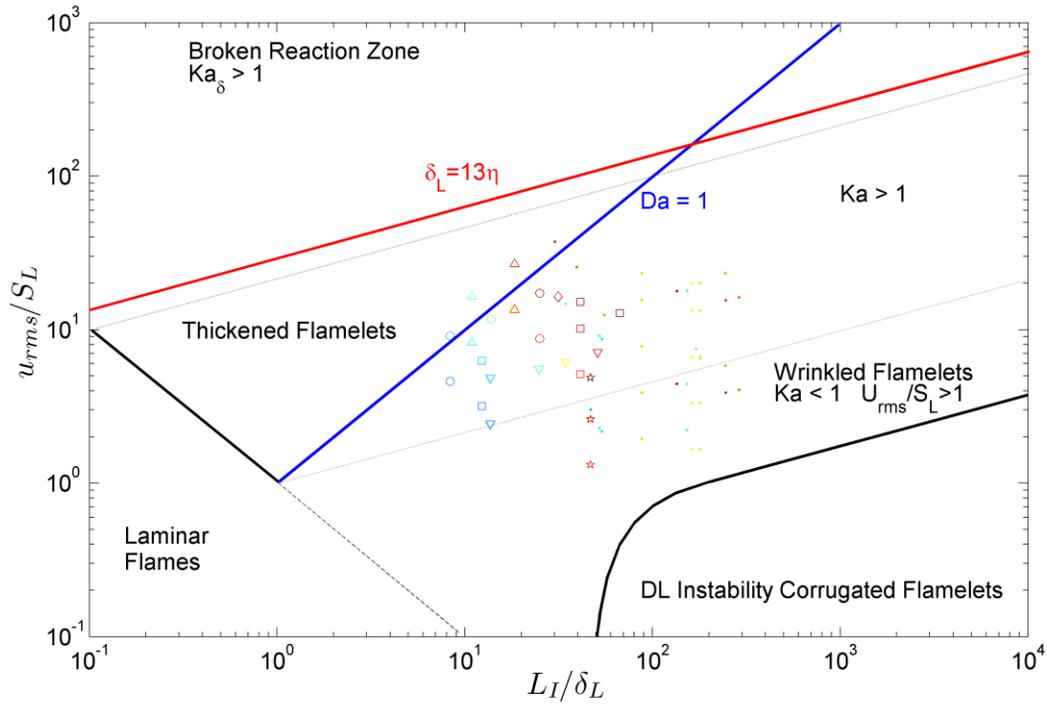

**Figure 3:** Regime Diagram with conditions of present ($C_0$-$C_4$) experiments and those for iso-octane data from [31]. The legends for present experimental conditions (symbols) and corresponding laminar flame and turbulence parameters could be found in Table 1. The legends (dots), flame and flow parameters for the Leeds data [31] could be found in Table 2.

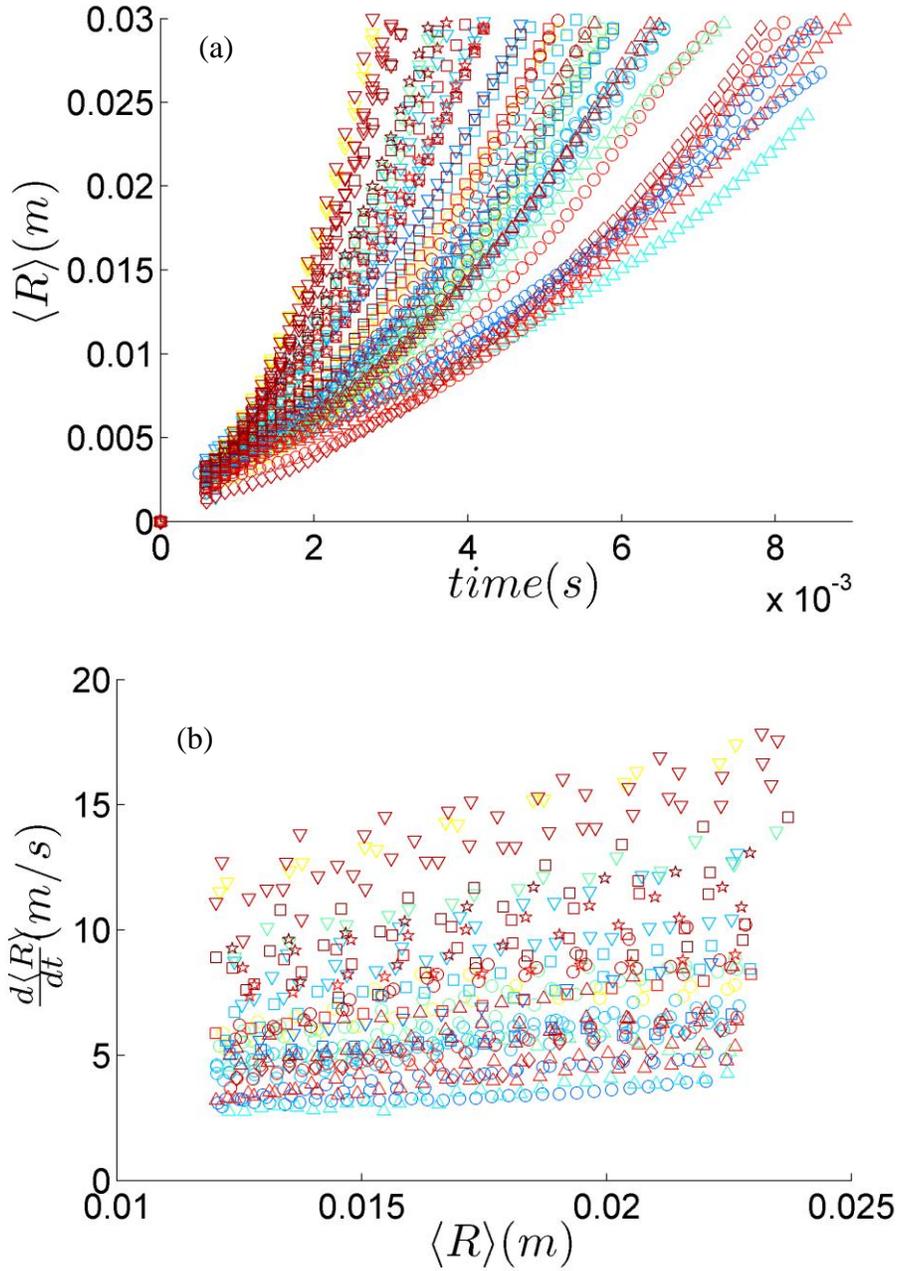

**Figure 4:**(a) $\langle R \rangle$ vs. time and (b) $d\langle R \rangle / dt$ vs. $\langle R \rangle$ for $H_2$/air, $\phi$=4.0; $CH_4$/air $\phi$=0.9; $C_2H_4$-15% $O_2$- 85% $N_2$, $\phi$=1.3; $C_2H_4$/air, $\phi$=1.3; n-$C_4H_{10}$/air, $\phi$=0.8 and $C_2H_6O$/air $\phi$=1.0 mixture from present experiments. The symbols for present experimental condition and corresponding laminar flame and turbulence parameters could be found in Table 1

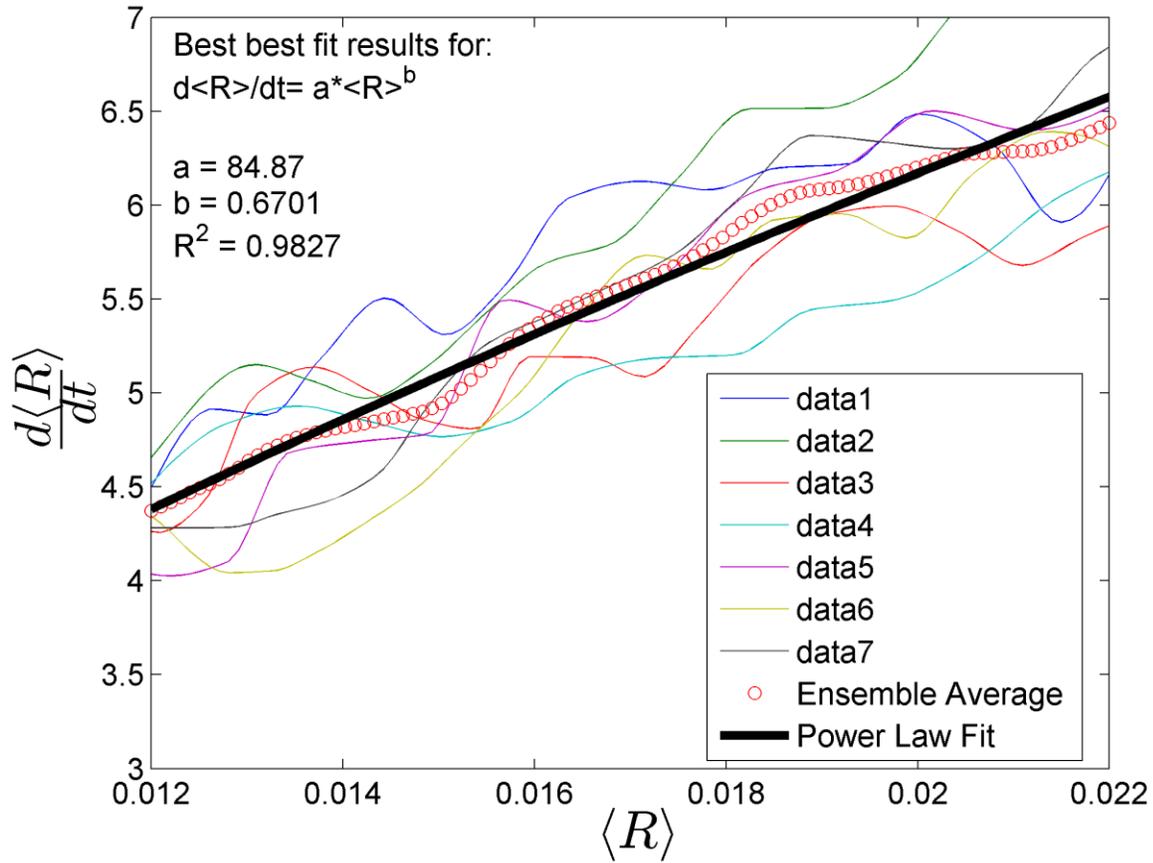

**Figure 5:** Individual $d\langle R\rangle/dt$ vs. $\langle R\rangle$ and their corresponding ensemble average for the CH$_4$/air, $\phi = 0.9$, $p$=1 atm, $u_{rms} \sim 2.85$m/s case. The inset text and the thick black line correspond to the best fit obtained by maximizing R$^2$.

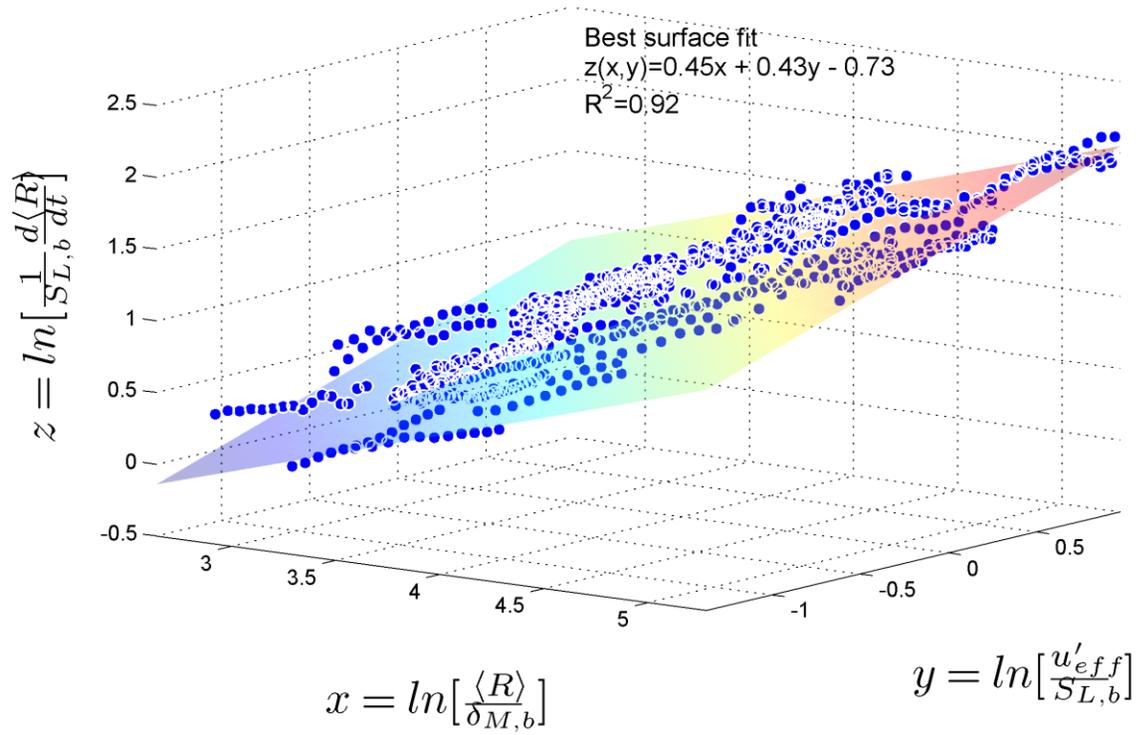

**Figure 6:** Best surface fit of the experimental data (C$_0$-C$_4$) for $\ln\left[\left(1/S_{L,b}\right)d\langle R\rangle/dt\right]$ as dependent variable with respect to $\ln(\langle R\rangle/\delta_{M,b})$ and $\ln(u'_{eff}/S_{L,b})$ as independent variables. The best fitting constants i.e. the power law exponents (maximizing R$^2$) are 0.45 and 0.43 respectively.

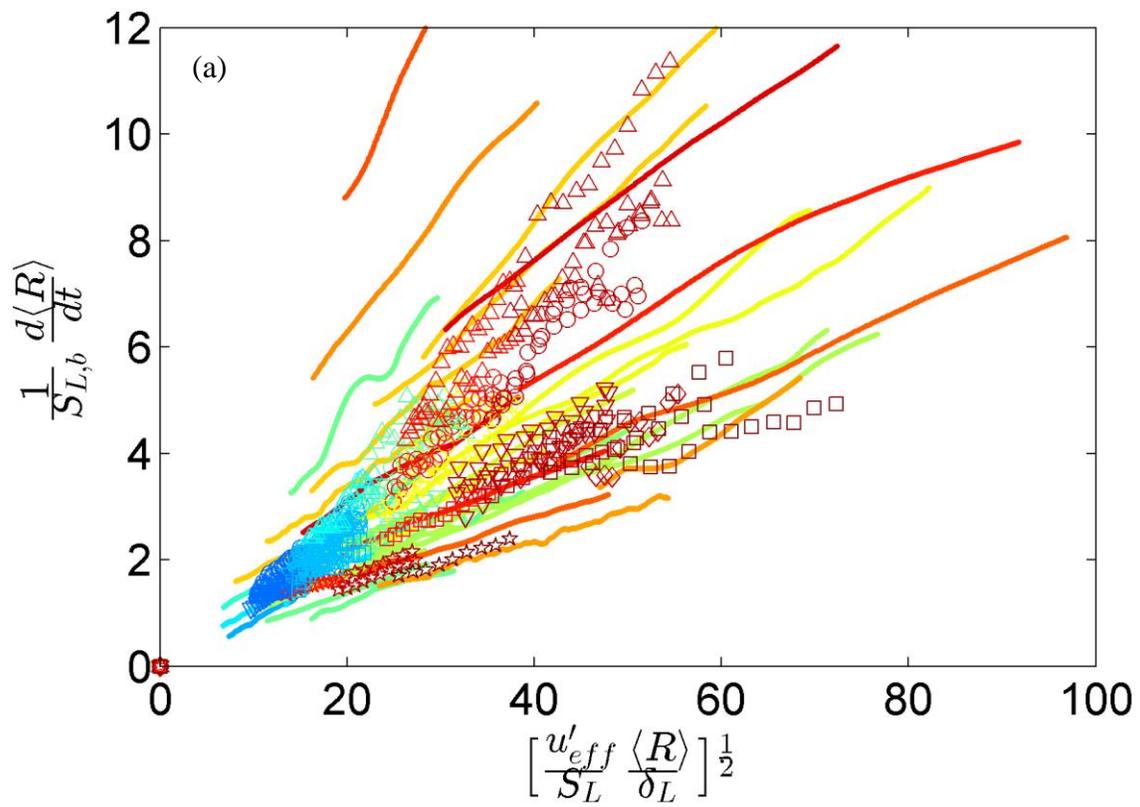

**Figure 7a**: (Caption after Figure 7c).

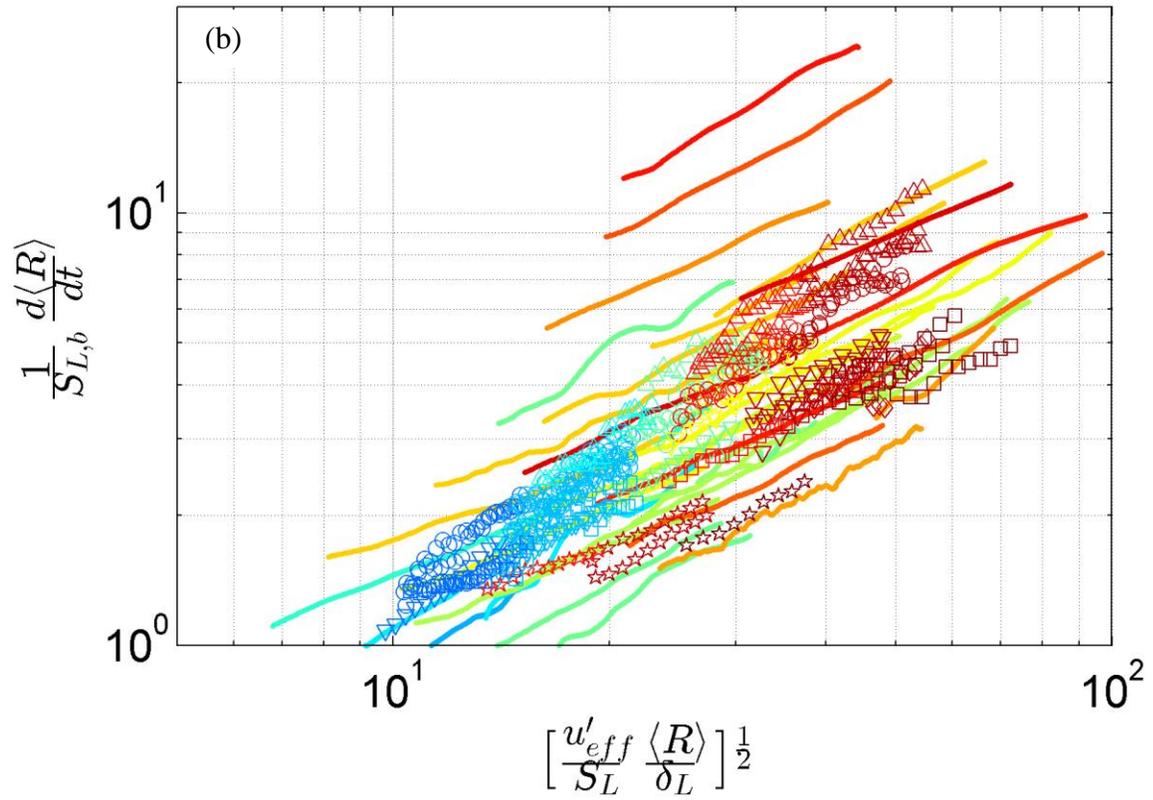

**Figure 7b:** (Caption after Figure 7c).

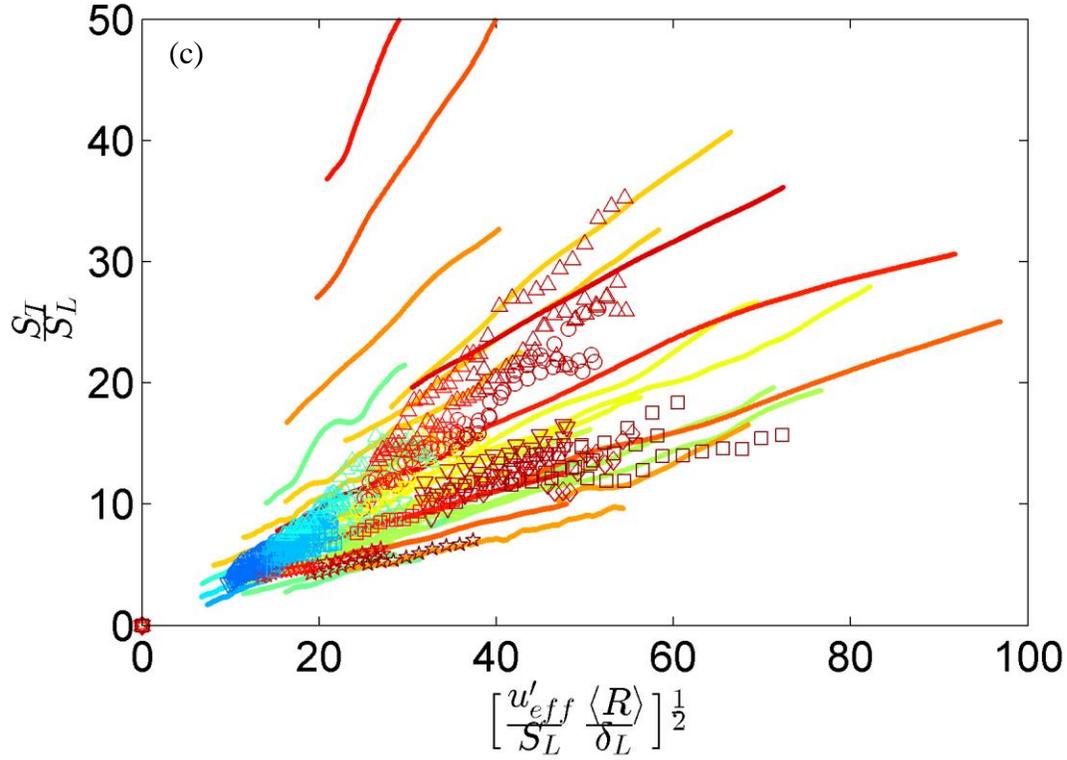

**Figure 7:** (a) Linear plot (b) Log-log plot of $S_{L,b}^{-1} d\langle R\rangle/dt$ (c) $S_{T,c=0.5}/S_L$ versus $\sqrt{(u'_{eff}/S_L)(\langle R\rangle/\delta_L)} = \text{Re}_{T,f}^{0.5}$ for H$_2$/air, $\phi=4.0$; CH$_4$/air $\phi=0.9$; C$_2$H$_4$-15% O$_2$- 85% N$_2$, $\phi=1.3$; C$_2$H$_4$/air, $\phi=1.3$; n-C$_4$H$_{10}$/air, $\phi=0.8$ and C$_2$H$_6$O/air $\phi=1.0$ mixture from present experiments. The lines represent the iso-C$_8$H$_{18}$ data from [31]. The symbols for present experimental condition and corresponding laminar flame and turbulence parameters could be found in Table 1. The symbols, flame and flow parameters for the Leeds data [31] could be found in Table 2. (a) Linear plot (b) Log-log plot.

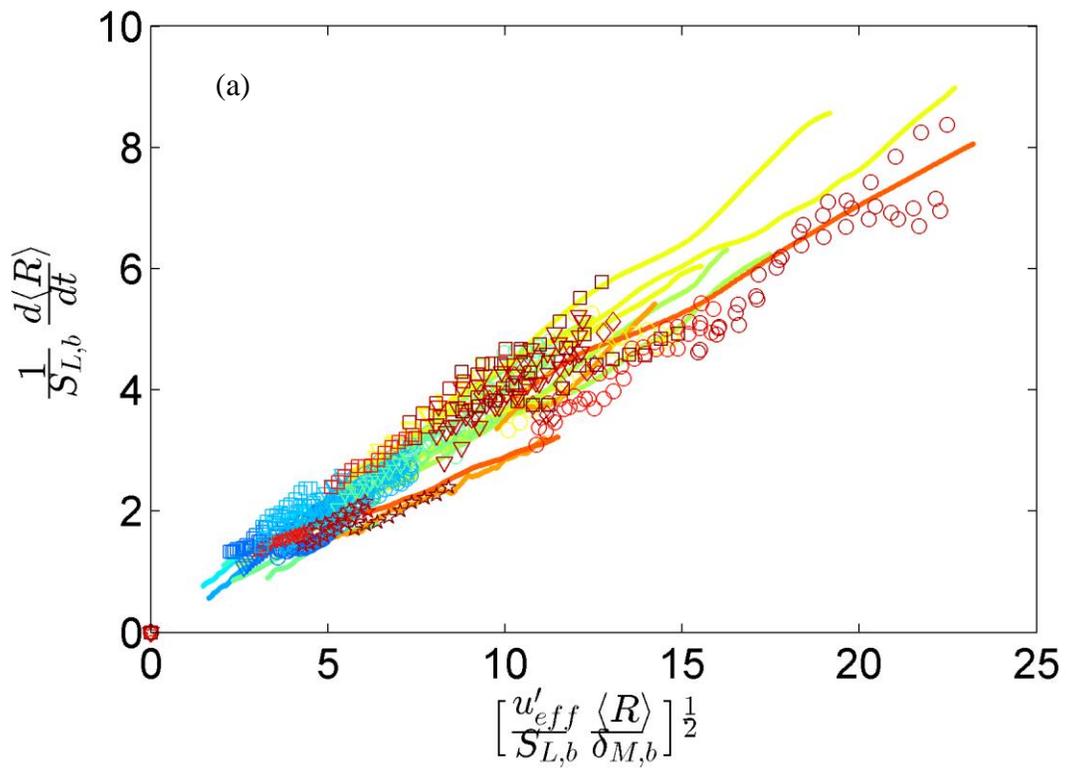

**Figure 8a:** (Caption after Figure 8d).

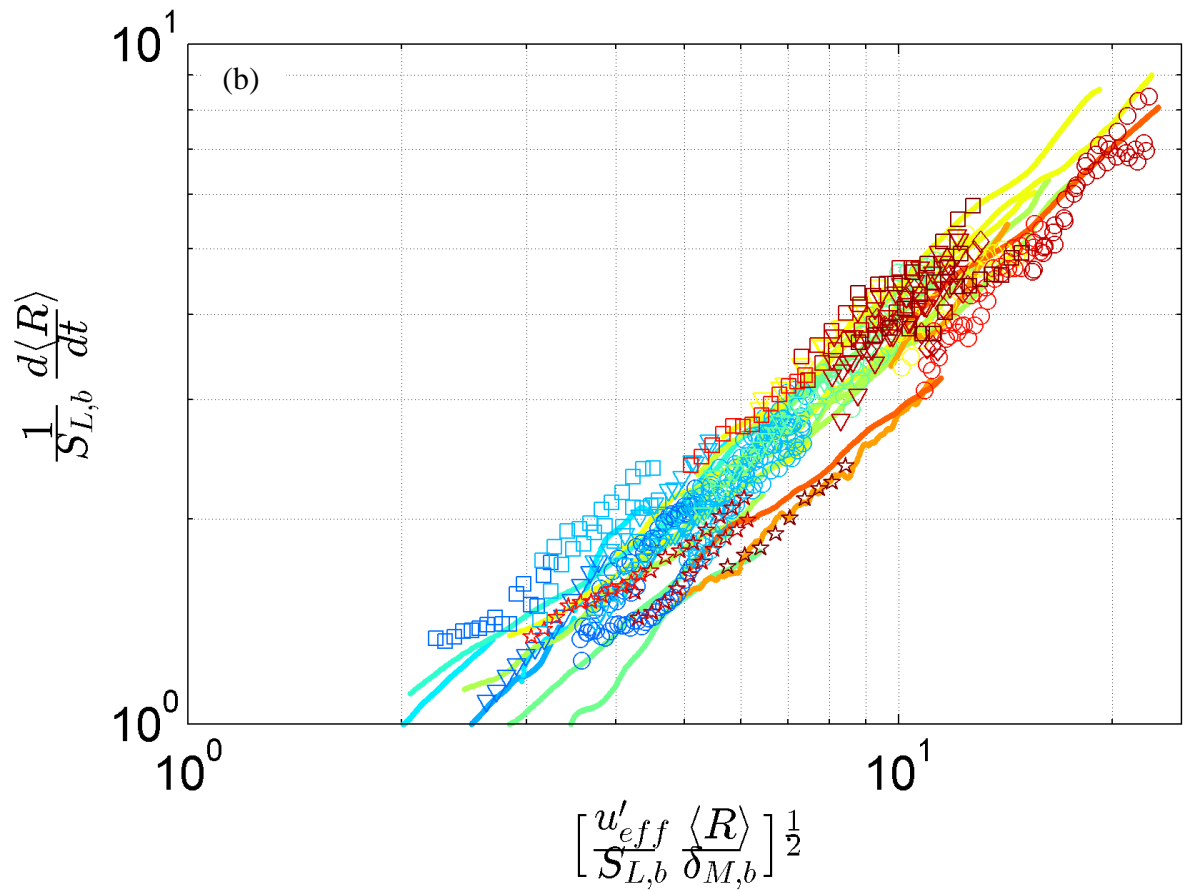

**Figure 8b:** (Caption after Figure 8d).

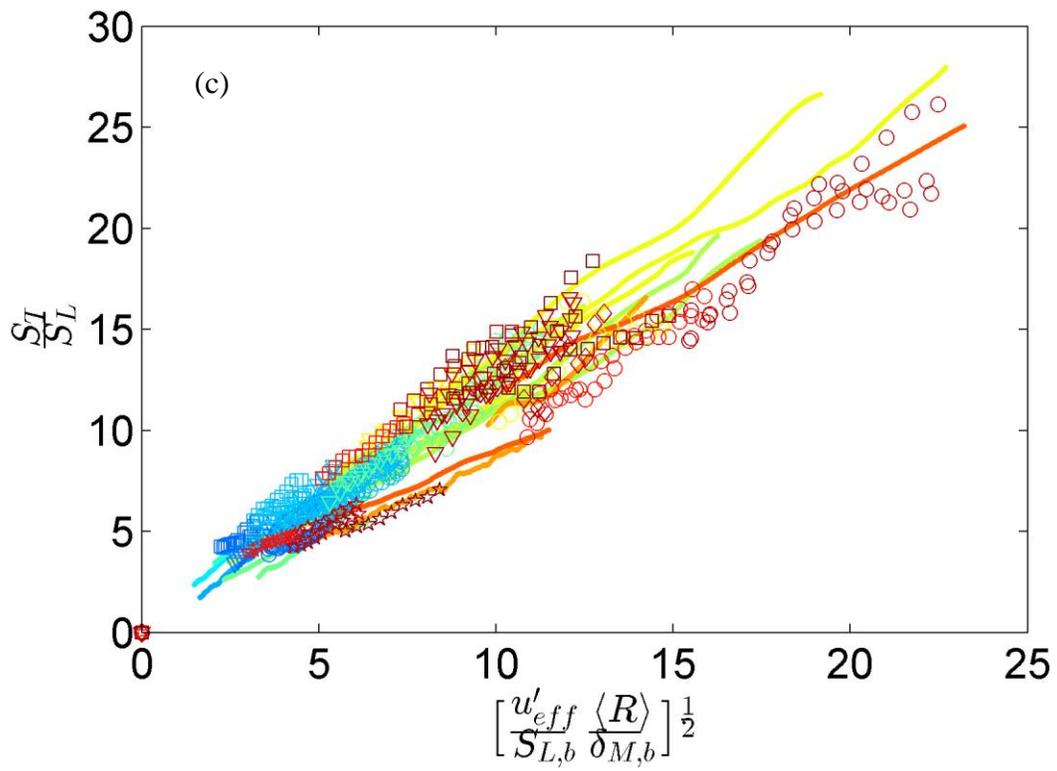

**Figure 8c:** (Caption after Figure 8d).

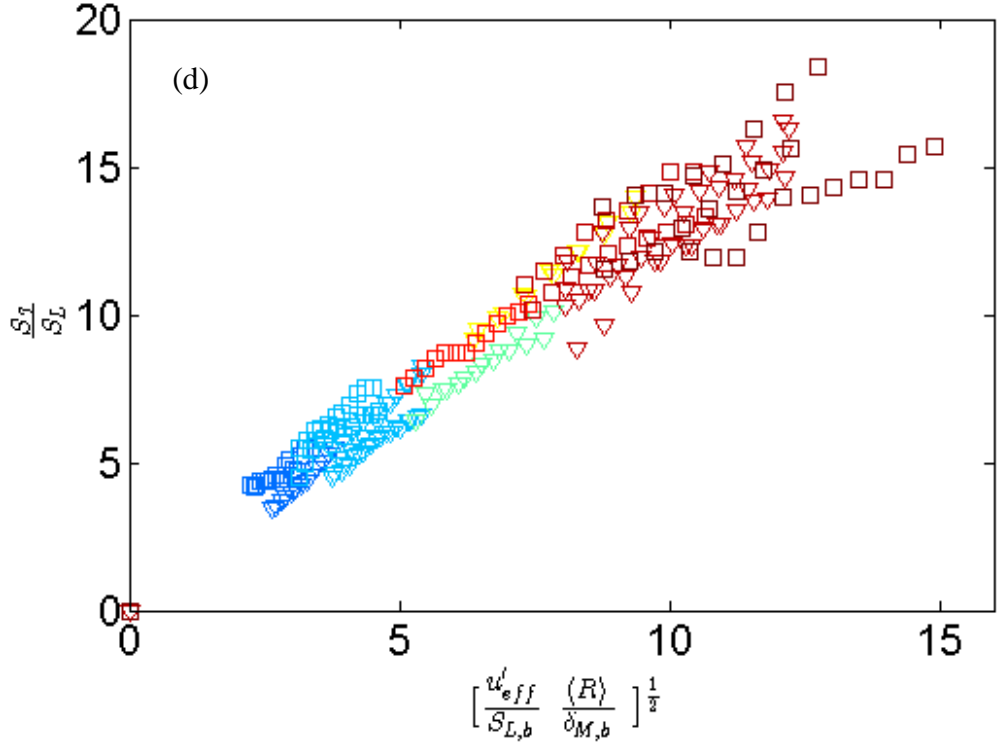

**Figure 8:** (a) Linear plot (b) Log-log plot of $S_{L,b}^{-1} d\langle R \rangle / dt$ (c) $S_{T,c=0.5}/S_L$ versus $\sqrt{(u'_{eff}/S_{L,b})(\langle R \rangle/\delta_{M,b})} = \text{Re}_{T,M}^{0.5}$ for H$_2$/air, $\phi$=4.0; CH$_4$/air $\phi$=0.9; C$_2$H$_4$/air, $\phi$=1.3; n-C$_4$H$_{10}$/air $\phi$=0.8; and C$_2$H$_6$O/air $\phi$=1.0 mixture from present experiments. The lines represent the iso-C$_8$H$_{18}$ data from [31]. The symbols for present experimental condition and corresponding laminar flame and turbulence parameters could be found in Table 1. The symbols, flame and flow parameters for the Leeds data [31] could be found in Table 2. (d) $S_{T,c=0.5}/S_L$ versus $\text{Re}_{T,M}^{0.5}$ for C$_2$H$_4$/air, $\phi$=1.3 (inverted triangles) and C$_2$H$_6$O/air $\phi$=1.0 (squares) mixture from present experiments. Only mixtures with $\delta_{M,b} > 0$ could be plotted in this figure.

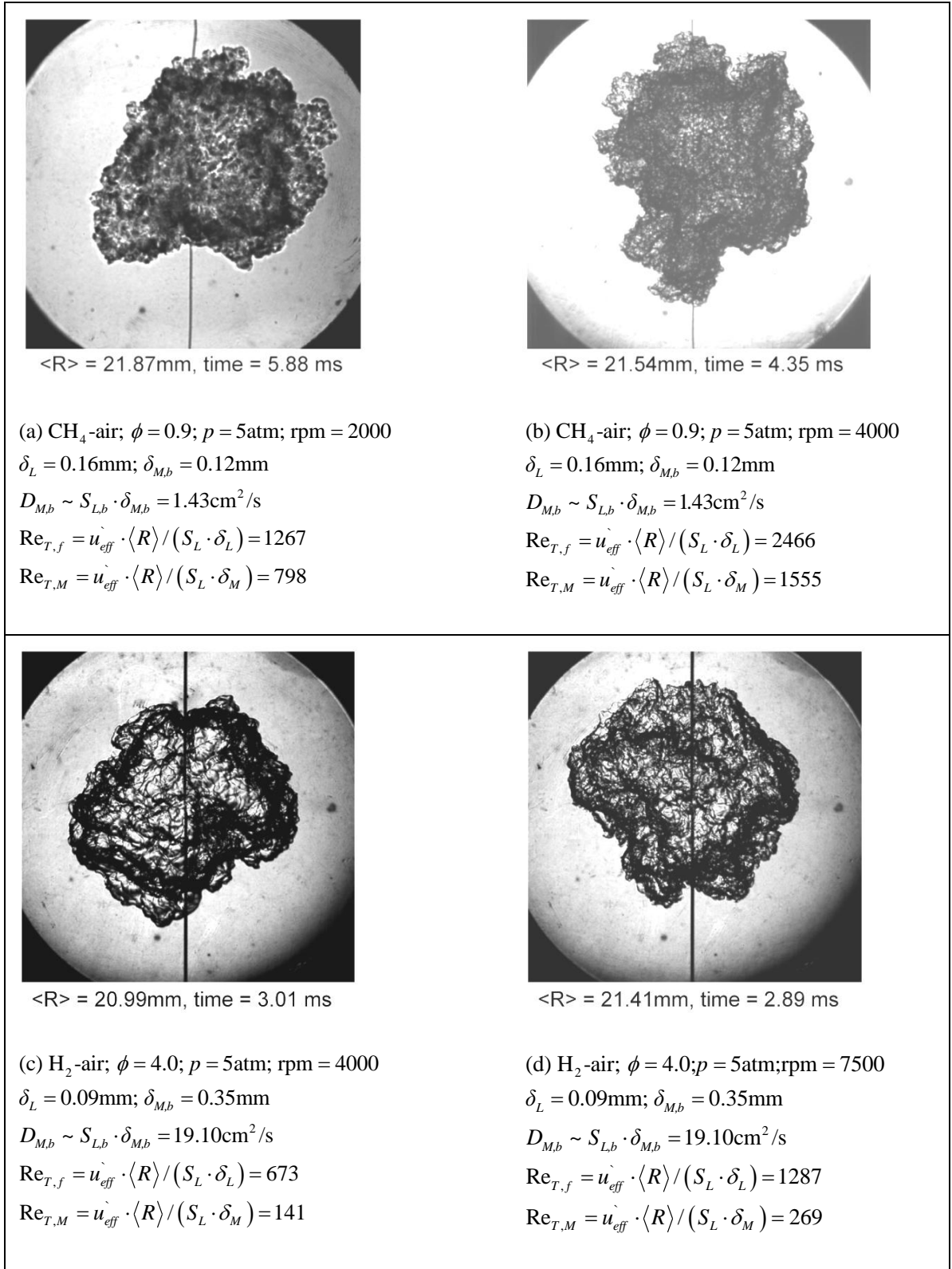

**Figure 9:** Effect of Markstein diffusivity on flame surface geomtery.

**Appendix Figures:**

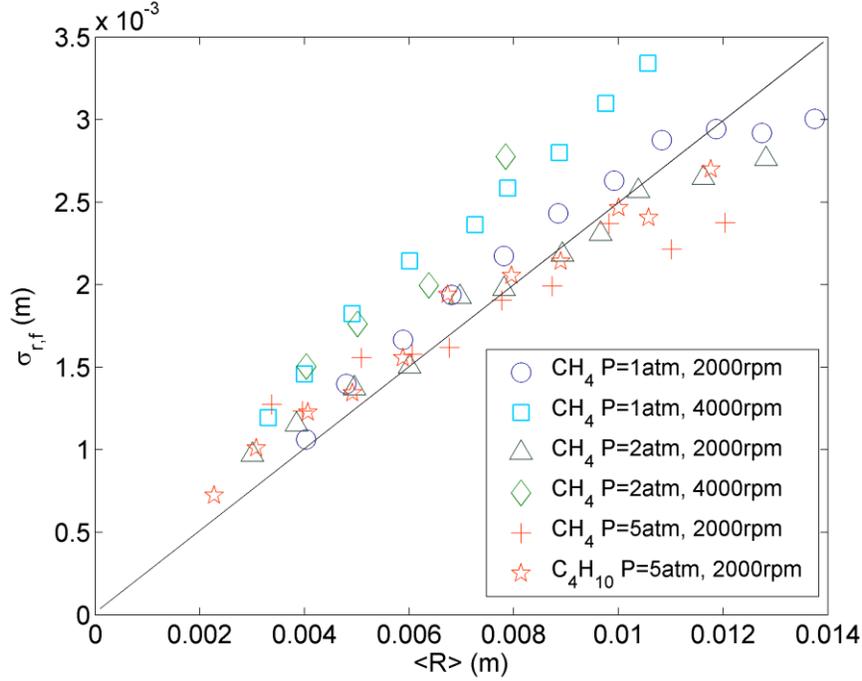

**Figure A1:** Standard deviation of flame radius obtained from Mie scattering images. $\sigma_{r,f}$ is obtained by taking the square root of the ensemble averaged variance of $R$ over at least six experimental runs, for each condition at each $\langle R \rangle$ location. This approximately suggests that $\delta_T \propto \sigma_{r,f} = \left\langle \left[ r_f - \langle r_f \rangle \right]^2 \right\rangle^{0.5} \propto \langle R \rangle$ within the limited range of $\langle R \rangle$ measurements. Since flame brush thickness is the length within which the entire flame is statistically contained, for a Gaussian distribution of $r_f$, a 96% probable confidence interval is given by $\delta_T \approx 4\sigma_{r,f}$. This yield $\delta_T \approx \langle R \rangle / 2$. The solid line is $4\sigma_{r,f} = \langle R \rangle / 2$. We note here that the $\langle R \rangle$ range measured here is smaller than that for the turbulent flame speeds, due to limitations of laser power/unit area at such high repetition rates. The measurements at largest $\langle R \rangle$ are more prone to lack of statistical convergence due to increasing standard deviation.

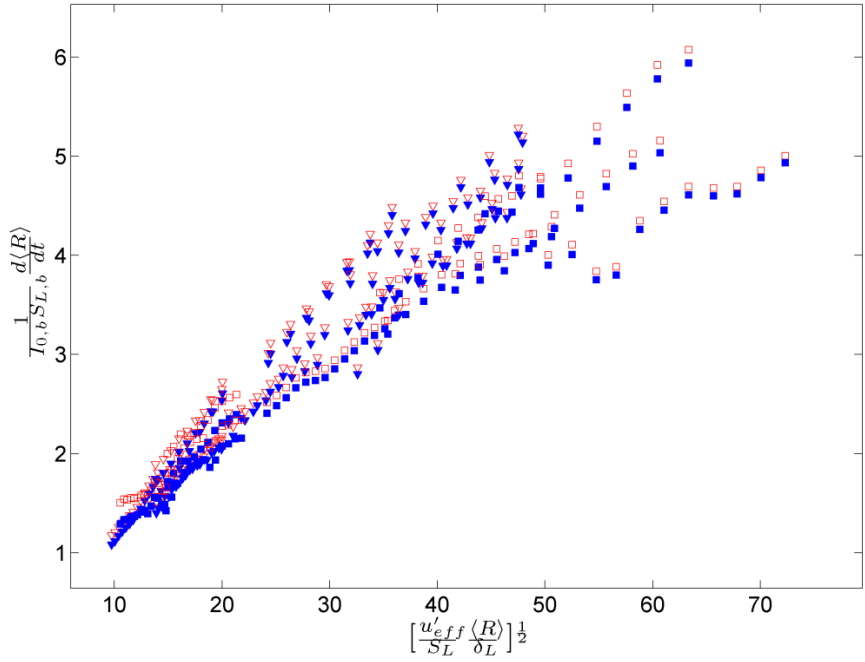

**Figure A2:** Effect of mean stretch due to curvature for the ethylene (inverted triangles) and DME (squares) data. The open symbols plot $\left(I_{0,b}S_{L,b}\right)^{-1}d\langle R\rangle/dt$ vs. $\left[\left(u'_{eff}/S_{L,b}\right)\left(\langle R\rangle/\delta_L\right)\right]^{1/2}$, while the closed symbols plot $\left(S_{L,b}\right)^{-1}d\langle R\rangle/dt$ vs. $\left[\left(u'_{eff}/S_{L,b}\right)\left(\langle R\rangle/\delta_L\right)\right]^{1/2}$ i.e. neglecting the effect of mean stretch due to mean curvature.

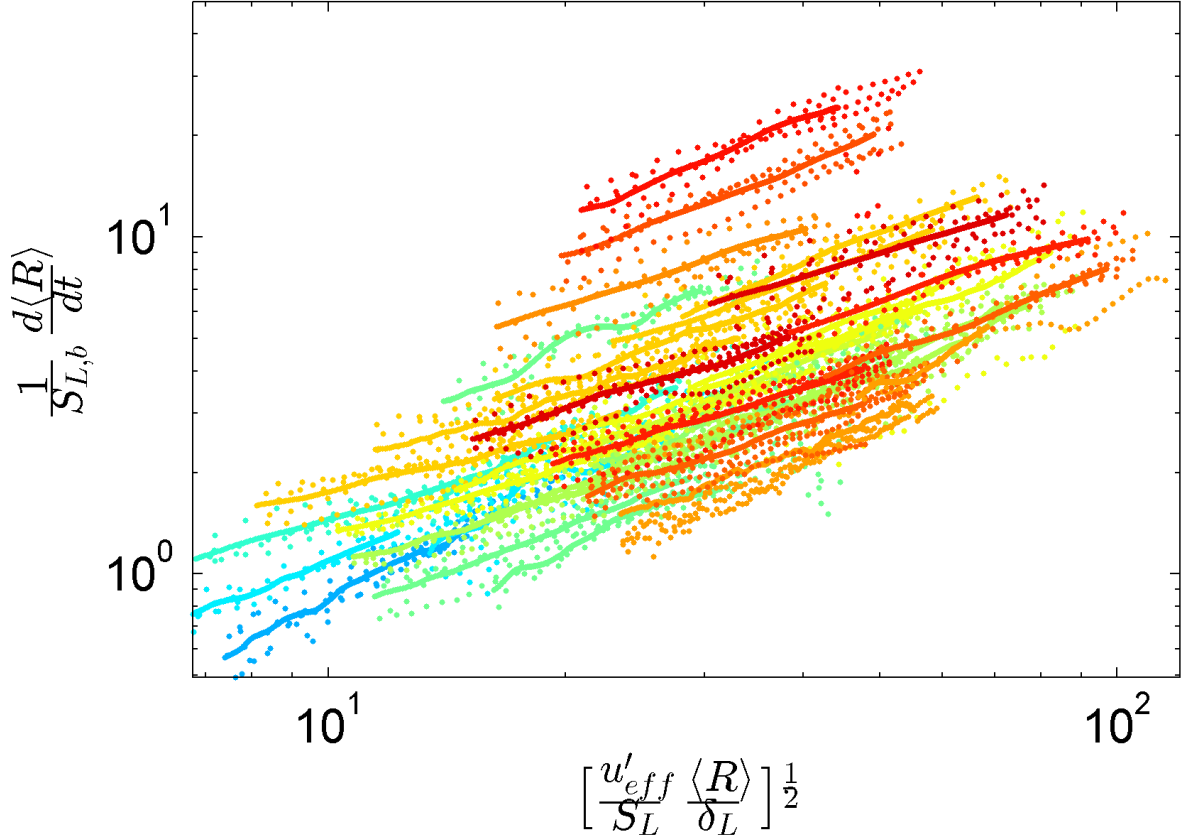

**Figure A3:** Instantaneous data from [31] superimposed with ensemble averaged data used in Figures 7-8. Here we assume $u'_{eff}(\langle R \rangle) \sim \langle R \rangle^{0.233}$ with the constant chosen such that at $\langle R \rangle = 0.04 m$, $u'_{eff}(\langle R \rangle) = u_{rms}$. Only the $u_{rms}$ values are reported in [31]. The fourth run in the $\phi = 1.0$, $p = 10$ atm, $u_{rms} = 4$ m/s was not considered due its quantitative discrepancy with the other five runs in the same dataset.